\documentclass[aps,prb,twocolumn,groupedaddress,showpacs]{revtex4}

\usepackage{graphicx}
\usepackage{latexsym}
\usepackage{amsmath,amssymb}
\usepackage{bm}
\DeclareMathOperator{\sgn}{sgn}

\begin{document}

\title{Interaction effects on 2D fermions with random hopping}

\author{Matthew S. Foster}
\email{psiborf@physics.ucsb.edu} 
\author{Andreas W. W. Ludwig}

\affiliation{Department of Physics, University of California, Santa
Barbara, CA 93106}

\date{\today}

\begin{abstract}
 We study the effects of generic short-ranged interactions on a system
of 2D Dirac fermions subject to a special kind of static disorder,
often referred to as ``chiral.'' The non-interacting system is a member
of the disorder class $B D$I [M.\ R.\ Zirnbauer, J.\ Math.\ Phys.\
\textbf{37}, 4986 (1996)]. It emerges, for example, as a low-energy
description of a time-reversal invariant tight-binding model of spinless 
fermions on a honeycomb lattice, subject to random hopping, and possessing particle-hole symmetry.
It is known that, in the absence of interactions, this disordered system is special in that it does not localize in 2D, but possesses extended
states and a finite conductivity at zero energy, as well as a strongly
divergent low-energy density of states.
In the context of the hopping model, the short-range interactions that 
we consider are particle-hole symmetric density-density interactions.
Using a perturbative one-loop renormalization group analysis, we show that the same mechanism responsible for the divergence of the density of states
in the non-interacting  system leads to an instability, in which the interactions are driven strongly relevant by the disorder. This result should be contrasted with the limit of clean Dirac fermions in 2D, which is stable against the inclusion of weak short-ranged interactions. Our work suggests a novel mechanism wherein a clean system, initially insensitive to interaction effects, can be made unstable to interactions upon the inclusion of weak static disorder. We dub this mechanism a ``disorder-driven Mott transition.'' Our result for 2D
fermions also contrasts sharply with known results in 1D, where a
similar delocalized phase has been shown to be robust against the
inclusion of weak interaction effects.  
\end{abstract}


\maketitle

\section{\label{Intro}INTRODUCTION}

The discovery\cite{ZIRNBAUER,AZ} of novel symmetry classes
of disordered quantum systems has reawakened interest in (``Anderson'')
localization physics. These classes describe the physics of non-interacting
quantum particles subject to static disorder potentials on scales much longer 
than the mean free path. Model realizations include descriptions of 
superconducting quasiparticles\cite{AZ,SFBN,SpinQH,SF,BSZ,VSF,RL,CRKHYL}, as 
well as models of electrons with random hopping on bipartite
lattices (``sublattice models'')\cite{GADE,FC,HWK,GLL,MDH2D,MRF}.
In contrast to the conventional disordered metals, characterized by the standard
and much studied (``unitary'', ``orthogonal,'' and ``symplectic'') Wigner-Dyson 
classes, models in these new classes often exhibit surprising low-dimensional
behavior. Some such disordered systems possess delocalized states in one and 
two dimensions, and these states can persist even in the presence of strong 
randomness. (See e.g.\ Refs.~[\onlinecite{GADE,GLL,DSF}].)

While much progress has been made in understanding the disorder physics
in such systems of non-interacting particles, the role of interparticle
interactions is often poorly understood. This issue is crucial to the
search for experimental realizations. In systems belonging to the
conventional (metallic) Wigner-Dyson classes, it is known\cite{BK} that
interaction effects can drastically alter the localization transitions.
For example, in all of these metallic universality classes, the incorporation 
of long-range Coulomb interactions destabilizes the non-interacting Anderson 
metal-insulator transition; in its place, a new, interacting ``Anderson-Mott''
transition is typically found in $D = 2 + \epsilon$ dimensions,\cite{BK}
characterized by distinct critical exponents. Renewed interest in these issues 
was recently generated through discussions concerning a metal-insulator
transition in 2D.\cite{2DMetalInsulTransition} 

In this paper, we examine the effects of generic short-ranged
interactions upon a 2D example of the so-called ``{\it sublattice}''
(or: {\it ``chiral''}) class denoted by $B D$I (in the classification scheme of
Ref.~\onlinecite{ZIRNBAUER}). As a particular representative of this class
we study a time-reversal invariant (TRI) model of spinless fermions
with random nearest-neighbor hopping on the honeycomb lattice. 
The random single-particle Hamiltonian $\bm{h}$ is special because the hopping
occurs \emph{only} between the two triangular sublattices of the honeycomb lattice
[Eq.(\ref{HLattice}) below]. This translates into an additional, discrete, 
so-called ``chiral'' symmetry, present in every realization of disorder
\begin{equation}\label{ChiralDef}
	-\sigma^{3} \bm{h} \sigma^{3} = \bm{h},
\end{equation} where $\sigma^3$ is a Pauli matrix diagonal in
sublattice space. The chiral symmetry can be thought of as a combination 
of time-reversal and particle-hole symmetries. The honeycomb model admits 
a low-energy description in terms of Dirac fermions subject to two types of 
long-wavelength random potentials: a pair of random masses, and a random vector
potential. The resulting disordered Dirac effective field theory also applies to 
the random bond $\pi$-flux model on the square lattice.\cite{FRADKIN,HWK,LFSG} 
The theory is described\cite{HWK,GLL} by two independent parameters, which we will call $g$ 
and $g_{A}$, characterizing the sample-to-sample fluctuations of the
zero mean Gaussian random mass and random vector potential disorders,
respectively (see Sec.~\ref{LatticeSym} for a review).

A key theoretical motivation for studying the disordered Dirac model with chiral 
symmetry is that it constitutes one of the \emph{simplest} known arenas
for investigating several non-trivial aspects of disorder-dominated
quantum criticality. While some features of the physics of the sublattice model 
are certainly tied to the special chiral symmetry, others, particularly the 
multifractal nature of the extended wavefunctions
\cite{LFSG,HWK,CMW,CCFGM,WEGNER,MIRLIN,EversMildenbergerMirlin} are
believed to be more general features of (Anderson) localization
critical points (see e.g.\ Refs.~[\onlinecite{WEGNER,MIRLIN}]).
Chiral disorder models exhibit the most interesting physics in one and two
dimensions, where these systems have proven amenable to a variety of
analytical techniques, and several exact and/or non-perturbative
results are known\cite{GLL,MDH2D,MRF,DSF,MDH1D,BOUCHAUD}.

Models in the chiral symmetry classes were originally studied 
analytically by Gade and Wegner\cite{GADE}, who formulated a non-linear sigma 
model (NL$\sigma$M) appropriate to a system of non-interacting fermions
subject to chiral disorder.\footnote{
Gade and Wegner considered models which had a Fermi surface in the clean limit.
On the other hand, the chiral class NL$\sigma$M for a system with a clean limit 
corresponding to Dirac fermions contains, in addition, a Wess-Zumino-Novikov-Witten
term. See e.g.\ Refs.\ \onlinecite{GLL} and \onlinecite{FK}.}
Gade and Wegner studied cases both with and without time-reversal invariance in 
the weak coupling limit and found essentially identical physics. Just as the Dirac 
theory mentioned above, the NL$\sigma$M also contains two parameters: the first is 
the usual dimensionless DC resistance, which is associated with the random mass
coupling $g$ in our Dirac theory. The parameter $g_{A}$, on the other hand, is 
associated with an additional ``U(1)-term'', which is present only in the chiral 
class NL$\sigma$Ms. A perturbative one-loop renormalization group (RG) analysis 
performed in Ref.\ \onlinecite{GADE} yielded that, in 2D, the DC resistance does not
renormalize, while the strength of the ``U(1)-term'' is driven to strong coupling.
Numerical work on 2D chiral models includes that listed e.g.\ in 
Refs.~\onlinecite{LeeFisher,Furusaki,RyuHatsugai,BocquetChalker,CastroNeto}. 

The TRI chiral Dirac theory studied in the present paper was introduced by Hatsugai, 
Wen, and Kohmoto in Ref.~\onlinecite{HWK}, where results of a one-loop RG analysis, 
adapted from Ref.~\onlinecite{BERNARD}, were presented. Using techniques from conformal 
field theory, Guruswamy, LeClair, and Ludwig\cite{GLL} later performed a 
non-perturbative analysis (in $g$ and $g_{A}$). In particular, closely parallel
to the results for the NL$\sigma$M discussed above, the strength $g$ of the
random mass is an exactly marginal perturbation (in the RG sense) to the clean Dirac 
theory, while the strength $g_{A}$ of the random vector potential is driven to strong 
coupling by the presence of a non-zero $g$. The pure marginality of $g$ indicates the 
existence of a critical, delocalized phase at zero energy, made possible by the
special chiral symmetry. The running $g_{A}$ parameter leads to the prediction of a 
strongly divergent low-energy density of states
\begin{equation}\label{DOS1}
	\nu(\omega) \sim \frac{1}{\omega} \exp\left(-c \, {\left| \ln \omega \right|}^{\alpha} \right),
\end{equation} 
where the scale-independent constant $c$ is a function of $g$ and the
exponent $\alpha = 1/2$.

More recently, Motrunich, Damle, and Huse\cite{MDH2D} used a strong
randomness RG picture to argue that, at asymptotically low energies, or
equivalently, large values of the parameter $g_A$, the density of
states of the chiral Dirac model should retain the form as in
Eq.~(\ref{DOS1}), but with a different exponent $\alpha = 2/3$. Mudry,
Ryu, and Furusaki\cite{MRF} observed that the freezing transition\cite{CMW,CCFGM},
which was known to occur for vanishing $g$ (the abelian gauge potential 
model\cite{LFSG,MudryChamonWen}) when $g_{A}$ exceeds some critical value 
$g^{\,c}_{A}$, affects the dynamic critical exponent $z$, and thus the low-energy 
density of states. They noted that the previous analytical RG treatments 
did not take into account an infinite set of operators associated with 
moments of the local density of states (LDOS). These operators carry increasingly 
negative (i.e.\ increasingly relevant) scaling dimensions, whose appearance
is taken to signal the broadening of the probability distribution function for the 
LDOS. Using a functional renormalization group treatment pioneered by Carpentier 
and Le Doussal\cite{CLD} to organize the diverging moments of the LDOS, the authors 
of Ref.~\onlinecite{MRF} derived the dynamic critical exponent analytically and 
thereby confirmed the asymptotic behavior of the global density of states with 
$\alpha =2/3$ of Ref.~\onlinecite{MDH2D}.

\begin{figure}
\includegraphics[width=0.3\textwidth]{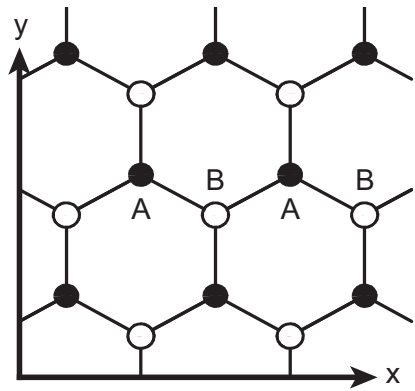}
\caption{The honeycomb lattice.\label{FigHoneycomb}}
\end{figure}

A second motivating factor for our work is that the 1D version of our
2D sublattice model studied in this paper is known to be stable to the 
inclusion of weak short-ranged interactions (which are irrelevant in the RG 
sense)\cite{DSF,MDH1D}. The former model also possesses a strongly-divergent 
low-energy density of states, with a critically delocalized state at zero
energy.\cite{BOUCHAUD} The robust nature of the 1D critical phase to
the inclusion of interaction effects is somewhat counterintuitive,
given the picture one has of a large number of states crammed into a
very narrow window of energy. In this paper, we will show that this
result does \emph{not} generalize to 2D.

In the present paper, we add to the continuum Dirac description of the
disordered honeycomb hopping model generic short-ranged interactions
that preserve the chiral symmetry, defined for the Dirac model
below in Eq.~(\ref{C}), and in the paragraph following that equation.  
We use the Schwinger-Keldysh method\cite{KELDYSH} to perform the ensemble average 
over realizations of the disorder. We then employ a one-loop renormalization 
group treatment to investigate the stability of the disordered, critically 
delocalized phase to interaction effects.  We find that the same mechanism 
responsible for driving the divergence of the low-energy density of states 
also feeds into same-sublattice (e.g.\ next-nearest-neighbor) interactions, 
producing an instability where the interaction strength grows to large values.
We show that this instability always occurs at low enough energies, regardless
of whether or not the (non-interacting) disordered system reaches the 
regime beyond the freezing transition discussed above. We speculate that 
this instability corresponds to the onset of some kind of Mott insulating phase, 
dominated by interactions rather than disorder.

Our results are both expected and surprising: expected, because of the
strongly divergent density of states and the potentially fragile nature
of the low-dimensional, critical disorder-only physics (i.e.\ in the absence 
of interactions); surprising, first because, 
as was already mentioned,
the opposite result obtains in 1D, and second because the
clean Dirac fixed point is robust against all short-ranged interaction
effects. If the observed instability ultimately drives the
disordered chiral symmetric model into an interaction-dominated, homogeneous 
Mott insulating phase, then we have discovered a somewhat curious route: a clean
system, initially insensitive to interaction effects, can be made unstable to 
interactions upon the inclusion of (a special kind of) weak static disorder. One
might dub this a ``disorder-driven Mott transition.''

The organization of this paper is as follows. In Sec.~{\ref{Setup}} we define the 
honeycomb lattice model and summarize the disordered low-energy effective Dirac 
theory. We then include short-ranged interactions, and define a set of coupling 
constants to encode the model physics. In Sec.~{\ref{1loop}}, we set up a one-loop 
renormalization group treatment of the full, disordered and interacting 
field theory. We re-derive the weak-coupling RG flow equations of the non-interacting, 
disorder-only model\cite{HWK,BERNARD,GLL} as a check of our methodology. We then compute 
the scaling dimensions of the various interaction operators; these scaling dimensions 
are new. In Sec.~{\ref{Results}}, we solve the combined RG flow equations for the 
disorder and interaction couplings, and we interpret the results. The reader less 
interested in calculational details may skip Sec.~\ref{1loop}, and proceed from 
the end of Sec.~\ref{Setup} directly to Sec.~\ref{Results}.

\section{\label{Setup}MODEL AND SETUP}

\subsection{\label{LatticeSym}Lattice model and continuum limit}

\begin{figure}
\includegraphics[width=0.3\textwidth]{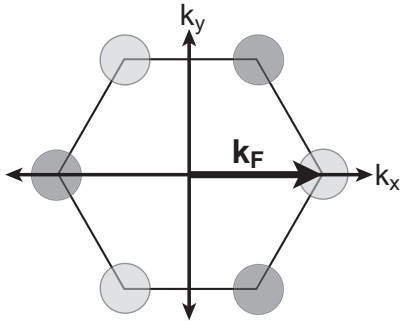}
\caption{Hexagonal Brillouin zone appropriate to the triangular sublattice. The independent zero energy Fermi points of the honeycomb model occur at $\pm \bm{k_{F}}$; low-energy states have wavenumbers $\bm{k}$ such that $\left|\bm{k} \mp \bm{k_{F}}\right| < \Lambda \ll k_{F}$.\label{FigHexBZ}}
\end{figure}

Consider a system of spinless fermions at half-filling on the honeycomb
lattice with nearest-neighbor hopping. The lattice is depicted in
Fig.~\ref{FigHoneycomb}. We define fermion annihilation operators $c_{A
i}$ and $c_{B j}$ on the ``A'' and ``B'' triangular sublattices,
respectively; then the random hopping Hamiltonian may be written as
 \begin{equation} \label{HLattice}
	H = \sum_{\langle {i j} \rangle} c_{A i}^{\dagger} \, t_{i j} \, c_{B j} + h.c., 
 \end{equation}
where the sum runs over all nearest-neighbor bonds on the honeycomb lattice. 
We take the hopping amplitudes $t_{i j}$ to be purely real. In the clean limit, 
$t_{i j} = t$; then the Hamiltonian in Eq.~(\ref{HLattice}) possesses two energy bands 
of Bloch states indexed over the hexagonal Brillouin zone (Fig.~\ref{FigHexBZ}), and, 
as in graphene, the low energy states reside near the corners of the zone 
where the two bands meet at isolated Fermi points.    

We truncate the Bloch spectrum of the clean system to low-energy modes
with wavenumbers $\bm{k}=(k_x,k_y)$ living within a momentum cutoff
$\Lambda$ of the zone corners. Alternate corners are equivalent by
reciprocal lattice translations, so we retain modes near two opposite
corners, which we index by the Fermi wavevectors $\pm \bm{k_F} = (\pm
k_F,0)$, shown in Fig.~\ref{FigHexBZ}. An effective low-energy Dirac
description emerges if we construct the 4-component
spinor
 \begin{equation} \label{PsiK}
	\psi(\bm{k}) \equiv \left[\begin{array}{c}  
			c_{A}(\bm{k}+\bm{k_F}) \\
			c_{B}(\bm{k}+\bm{k_F}) \\
			c_{A}(\bm{k}-\bm{k_F}) \\
			c_{B}(\bm{k}-\bm{k_F})
			\end{array}	
			\right],
 \end{equation} 
with $\left| \bm{k} \right| < \Lambda$. Linearizing the
energy bands near the Fermi points and incorporating the
hopping disorder as a perturbation, we find the non-interacting 
Hamiltonian
 \begin{eqnarray}
	H_{D} & = & - \int d^2r \, \psi^{\dagger}(\bm{r}) \left[v_{F} \, {\alpha_{\mu} \left( i\,\partial_{\mu} + A_{\mu} \, \beta^{0} \right)}\right] \psi(\bm{r}) \nonumber \quad \\
	& & + \int d^2r \, \psi^{\dagger}(\bm{r}) \left[m^{i}
\beta^{i} \,\right] \psi(\bm{r}),\quad \label{HNonInt}
 \end{eqnarray}
where $v_F$ is the Fermi velocity, and $\psi(\bm{r})$ is the
long-wavelength Fourier transform of (\ref{PsiK}). The disorder emerges
in the continuum theory (\ref{HNonInt}) as a random vector potential
$A_{\mu}(\bm{x})$, $\mu \in \{x,y\}$, and a pair of random Dirac masses
$m^{i}(\bm{x})$, $i \in \{1,2\}$. In Eq.~(\ref{HNonInt}), summation on
repeated indices is implied and will be assumed throughout the rest of
the paper.

The symbols $(\alpha_{x},\alpha_{y})$ and
$\{\beta^{0},\beta^{1},\beta^{2}\}$ in Eq.~(\ref{HNonInt}) denote
various 4$\times$4 matrices. Explicit representations can be derived
from the honeycomb lattice model. We introduce two sets of 
Pauli matrices:\footnote{We employ the conventional 
basis for all Pauli matrices.}
the matrix $\sigma^{\lambda}$ acts on the sublattice (A-B)
space, while the matrix $\tau^{\gamma}$ acts on the Fermi point
``flavor'' space, with $\lambda,\gamma \in \{1,2,3\}$.
Then
 \begin{subequations}
\label{MatrixDef}
\begin{align}
	(\alpha_{x},\alpha_{y}) & = (-\sigma^{1} \tau^{3},-\sigma^{2}), \\
\intertext{and}
	\{\beta^{0},\beta^{1},\beta^{2}\} & = \{\tau^{3} , \sigma^{1} \tau^{1} , \sigma^{1} \tau^{2} \}.
\end{align}
\end{subequations}

The disordered Dirac Hamiltonian $H_{D}$ given by equation
(\ref{HNonInt}) is the most general non-interacting 4-component Dirac
Hamiltonian (up to irrelevant terms) consistent with the defining
symmetries of the chiral sublattice model. The disordered lattice
Hamiltonian defined in Eq.~(\ref{HLattice}) is invariant under
time-reversal ($\mathcal{T}$) and chiral ($\mathcal{C}$)
transformations. In the continuum theory, these transformations appear
as
 \begin{equation}\label{T}
	\mathcal{T}:\, \Psi \rightarrow \tau^{1} \Psi,\, i \rightarrow -i
 \end{equation}
and 
\begin{equation}\label{C}
	\mathcal{C}:\, \Psi \rightarrow \sigma^{3} (\Psi^{\dagger})^{T},\, i \rightarrow -i
\end{equation}
 where $T$ denotes matrix transpose. Note that \emph{both} the
time-reversal and chiral transformations are antiunitary in this
language. We stress that in the $2^{nd}$-quantized formulation discussed here,
the ``chiral symmetry'' is represented by the fact that the 
(second-quantized) Dirac Hamiltonian, given by Eq.~(\ref{HNonInt}),
is \emph{invariant} under the transformation defined by Eq.~(\ref{C}). 
This is easily seen to be equivalent to the \emph{first-quantized} definition in 
Eq.~(\ref{ChiralDef}), above.---The Hamiltonian $H_{D}$ in Eq.~(\ref{HNonInt}) 
was originally derived in the context of the random bond $\pi$-flux model on the
square lattice;\cite{HWK,FRADKIN,LFSG} note that on the honeycomb lattice, we 
obtain the same effective field theory in zero magnetic flux.

We take $A_{\mu}(\bm{x})$ and $m^{i}(\bm{x})$ to be Gaussian
white-noise--distributed random variables, and we require that the
ensemble-averaged system be invariant under honeycomb lattice
rotations, translations, and reflections. These assumptions lead to the
conditions
 \begin{equation}\label{NoAvgAandM}
	\overline{A_{\mu}(\bm{x})} = \overline{m^{i}(\bm{x})} = 0,
 \end{equation} 
\begin{equation}\label{gADef}
	 \overline{A_{\mu}(\bm{x})\,A_{\nu}(\bm{x'})} = g_{A} \, 2 \pi \, \delta_{\mu \nu} \delta(\bm{x} - \bm{x'}),
\end{equation} 
and
\begin{equation}\label{gDef}
	\overline{m^{i}(\bm{x})\,m^{j}(\bm{x'})} = g \, 2 \pi \, \delta^{i j} \delta(\bm{x} - \bm{x'}). 
\end{equation}
 In these equations the overbar denotes disorder averaging, and the
factors of $2 \pi$ are conventional. The disorder-averaged theory is
characterized by two parameters, the 
variances $g$ and $g_{A}$
of the random mass and random vector potential, respectively.

An interpretation of the two types of disorder appearing in
Eq.~(\ref{HNonInt}) in terms of the physics of the honeycomb lattice
model is obtained by examining the structures of the matrices in
Eq.~(\ref{MatrixDef}), used to define them. The random masses couple
the two Fermi nodes together in flavor space, while the random vector
potential does not. Thus the random mass parameter $g$ characterizes
elastic backscattering between the two independent Fermi nodes
(involving large crystal momentum transfers), while the random vector
potential parameter $g_{A}$ characterizes intranode scattering
involving small crystal momentum transfers. As discussed in
Sec.~\ref{Intro}, $g$ turns out to be related to the dimensionless DC
resistance, which is \emph{independent} of $g_{A}$;
on the other hand, $g_{A}$ dominates ``one-particle'' properties, including
the density of states.\cite{GLL} This separation of one and
two-particle properties is a characteristic feature of the chiral
disorder classes.

\subsection{Adding interactions}\label{Interactions}

We complete the setup of our problem in this subsection by adding
to the disordered Dirac model [defined by Eq.~(\ref{HNonInt})] generic 
short-ranged interactions that preserve the chiral (sublattice) symmetry. 
Consider the set of all four-fermion operators in
the low-energy effective continuum theory that are purely local in
space and time, and involve no derivatives. All such operators are
irrelevant at the clean Dirac fixed point, but comprise the
\emph{least} irrelevant set of local multiparticle interaction terms
that we can write down. One then expects that these should be the most
important local operators to study once we turn on the disorder. We can
enumerate all independent four-fermion terms consistent with (i) the
discrete $\mathcal{T}$ and $\mathcal{C}$ symmetries [Eqs.~(\ref{T}) and
(\ref{C})], and with (ii) the assumptions of statistical invariance
under honeycomb rotation, translation, and reflection transformations.
It turns out that there are four such local operators; in this
subsection, however, we will only discuss two of them in detail. (The
other two turn out to be even more irrelevant in the disordered phase
than at the clean Dirac fixed point. In Sec.~\ref{1loop}, below, we
will calculate the anomalous dimensions of all four operators.)

Consider the local fermion density operators $n_{A i} = c_{A
i}^{\dagger} c_{A i}$ and $n_{B j} = c_{B j}^{\dagger} c_{B j}$ defined
on the A and B sublattices of the honeycomb model. In the continuum
Dirac theory, we define
 \begin{subequations} \label{LocalDensity}
 \begin{equation} 
	\bar{\rho}(\bm{r}) \equiv \frac{1}{2} \psi^{\dagger}(\bm{r}) \left( 1 + \sigma^3 \right) \psi(\bm{r}) \sim n_{A}(\bm{r})
 \end{equation}
\text{and}
\begin{equation}
	\rho(\bm{r}) \equiv \frac{1}{2} \psi^{\dagger}(\bm{r}) \left( 1 - \sigma^3 \right) \psi(\bm{r}) \sim n_{B}(\bm{r}),
 \end{equation}
 \end{subequations}
where the identification with the sublattice operators holds up to
terms that oscillate on the lattice scale.

The two important four-fermion interactions may be written simply in
terms of products of the long-wavelength sublattice density operators
defined in Eq.~(\ref{LocalDensity}). They are $\bar{\rho} \rho(\bm{r})$
on one hand, and $(\bar{\rho}^{2} + {\rho}^{2})(\bm{r})$ on the other.
The operator $\bar{\rho} \rho$, which involves a product of the
long-wavelength densities from the two different sublattices, can be
shown to result from the continuum limit of, for example,
nearest-neighbor density-density interactions:
\footnote{
Terms bilinear in fermions are forbidden
in the path integral formulation [Eqs.~(\ref{ZFull}-\ref{SInt})] 
on the r.h.s. of Eq.~(\ref{RhoBarRhoNN}) by the
discrete $\mathcal{T}$ and
$\mathcal{C}$ symmetries [Eqs.~(\ref{T}) and (\ref{C})].
}
\begin{equation}\label{RhoBarRhoNN}
	\sum_{\langle i j \rangle} \left( n_{A i} -\textstyle{\frac{1}{2}} \right)\!\left( n_{B j} -\textstyle{\frac{1}{2}} \right)
	\sim \int d^{2}r \, \bar{\rho} \rho (\bm{r}).
\end{equation}   

The other important four-fermion local operator is $\bar{\rho}^{2} +
{\rho}^{2}$.
Microscopically, the operators $\bar{\rho}^{2}$ and $\rho^{2}$ are pure
intrasublattice interactions, i.e.\ they do not couple together
particles on the two different sublattices. These are the only such
four-fermion terms that we can write down. Note that invariance under
y-axis reflections (which exchange the two sublattices---see
Fig.~\ref{FigHoneycomb}) requires $\bar{\rho}^{2}$ and $\rho^{2}$ to
appear in the combination $\bar{\rho}^{2} + \rho^{2}$ with identical
coefficients. Thus $\bar{\rho}^{2} + {\rho}^{2}$ represents the most
relevant piece of arbitrary short-ranged same-sublattice (e.g.\
next-nearest-neighbor) interactions.

In order
to perform the average over disorder realizations, we
write a standard $(2+1)$-dimensional  (real-)time-ordered
($T$-ordered) Grassmann path integral $Z_{T}$ to represent the 
disordered Dirac theory. We use the Schwinger-Keldysh method\cite{KELDYSH} 
to normalize this path integral to unity in order to perform the disorder average; 
in this method, we exploit the fact that the normalization factor $1/Z_{T}$ can
be written simply as an anti--time-ordered ($\bar{T}$-ordered) path
integral $Z_{\bar{T}}$ for the same theory. This construction works
regardless of whether the model includes interparticle interactions or
not.

\begingroup
\squeezetable
\begin{table*}[bottom]
\caption{\label{ChiralSymTable} Continuum versions of honeycomb lattice model symmetry transformations in the L/R language.}
\begin{ruledtabular}
\begin{tabular}{l@{\qquad\qquad}l@{\,}l@{\,}l@{\,}l@{\quad}l}
Symmetry & \multicolumn{5}{c}{Expression in terms of 
left(L)- and right(R)-moving components} \\
\hline 
Chiral ($\mathcal{C}$) & $R(t) \rightarrow -\mu^{1} L(-t)$ & $L(t) \rightarrow -\mu^{1} R(-t)$ & $\bar{R}(t) \rightarrow \bar{L}(-t) \mu^{1}$ & $\bar{L}(t) \rightarrow \bar{R}(-t) \mu^{1}$ & $S \rightarrow S^{*}$\\
Time-reversal ($\mathcal{T}$) & $R(t) \rightarrow -L(-t)$ & $L(t) \rightarrow -R(-t)$ & $\bar{R}(t) \rightarrow \bar{L}(-t)$ & $\bar{L}(t) \rightarrow \bar{R}(-t)$ & $S \rightarrow S^{*}$ \\
x-Reflection\footnotemark[1] & $R \rightarrow -L$ & $L \rightarrow -R$ & $\bar{R} \rightarrow \bar{L}$ & $\bar{L} \rightarrow \bar{R}$ & \\
y-Reflection\footnotemark[1] & $R \rightarrow -i \mu^{2} \bar{L}^{T}$ & $L \rightarrow i \mu^{2} \bar{R}^{T}$ & $\bar{R} \rightarrow L^{T} i \mu^{2}$ & $\bar{L} \rightarrow -R^{T}i \mu^{2}$ & \\
Rotations\footnotemark[1] & $R \rightarrow e^{i \theta} R$ & $L \rightarrow e^{-i \theta} L$ & $\bar{R} \rightarrow \bar{R} e^{i \theta}$ & $\bar{L} \rightarrow \bar{L} e^{-i \theta}$ & \\
Translations\footnotemark[1] & $R \rightarrow e^{i \theta} R$ & $L \rightarrow e^{-i \theta} L$ & $\bar{R} \rightarrow \bar{R} e^{-i \theta}$ & $\bar{L} \rightarrow \bar{L} e^{i \theta}$ & \\
Electrical charge U(1) & $R \rightarrow e^{i \theta \mu^{3}} R$ & $L \rightarrow e^{i \theta \mu^{3}} L$ & $\bar{R} \rightarrow \bar{R} e^{-i \theta \mu^{3}}$ & $\bar{L} \rightarrow \bar{L} e^{-i \theta \mu^{3}}$ &
\end{tabular}
\end{ruledtabular}
\footnotetext[1]{Implicit transformation of the 
field argument $\bm{r}$ is implied.
$S=S_D + S_I$
denotes the action in Eq.(\ref{ZFull}).
}
\end{table*}
\endgroup

Specifically, we write the theory in terms of the Schwinger-Keldysh generating 
functional $Z$ defined as (see e.g.\ Ref.~\onlinecite{KELDYSH}) 
 \begin{equation}\label{ZFull}
	Z \equiv Z_{T} Z_{\bar{T}} = \int \mathcal{D}\psi_{a} \,\mathcal{D}\bar{\psi}_{a}\, e^{i \left( S_{D} + S_{I} \right) },
 \end{equation} where the non-interacting action $S_{D}$,
containing the disorder potentials,
is given by
 \begin{widetext}
\begin{equation}\label{SNonInt}
	S_{D} = \sum_{a=1,2} s_{a} \int d\omega \, d^{2}r \; \bar{\psi}_{a}(\omega,\bm{r})  \left[h \, \omega + i \, s_{a} \, \eta \, \sgn(\omega) + {\alpha_{\mu} \left( i\,\partial_{\mu} + A_{\mu} \, \beta^{0} \right) - m^{i} \beta^{i} \, } \right]  \psi_{a}(\omega,\bm{r}),  
 \end{equation}
while $S_{I}$ contains the four-fermion interactions
\begin{equation}\label{SInt}
	S_{I}=-\sum_{a=1,2} s_{a} \int dt \, d^{2}r \left[ U \left(\bar{\rho}_{a}^{2} + {\rho}_{a}^{2} \right)\!(t,\bm{r}) + W \bar{\rho}_{a} \rho_{a}(t,\bm{r}) + X \mathcal{O}^{X}_{a}(t,\bm{r}) + Y \mathcal{O}^{Y}_{a}(t,\bm{r})\right].
\end{equation}
\end{widetext}

The integration in Eq.~(\ref{ZFull}) is over the Grassmann fields
$\psi_{a}$ and $\bar{\psi}_{a}$, where the
(``Keldysh-'')
index $a \in \{1,2\}$ denotes the $T$-ordered ($a=1$) or
$\bar{T}$-ordered ($a=2$) copy of the theory. The factor $s_{a}$
appearing in the action $S_{D}$ (\ref{SNonInt}) is given by
 \begin{equation}\label{KeldyshFactor}
	s_{a} = \left\{ \begin{array}{lll}
			\phantom{-}1,\; & a=1	& (T\text{-ordered}), \\
			-1,\; 		& a=2 	& (\bar{T}\text{-ordered}).
			\end{array} \right.
 \end{equation}
The term $i \, s_{a} \, \eta \, \sgn(\omega)$ in $S_{D}$ is the
pole-prescription appropriate to $T$ and $\bar{T}$-ordered Green's
functions, with $\eta~\rightarrow~0^{+}$. In Eq.~(\ref{SNonInt}), we
have set the Fermi velocity $v_{F} = 1$, and the various 4$\times$4
matrices appearing in the theory 
were defined in Eq.~(\ref{MatrixDef}). Finally, the coupling
constant $h$ in Eq.~({\ref{SNonInt}) is a dynamic scale factor, and will be used to track the
relationship between length and energy as we renormalize the theory.

In Eq.~(\ref{SInt}), we have introduced coupling constants $U$ and $W$
for the $\bar{\rho}^{2}+ \rho^{2}$ and $\bar{\rho} \rho$ interaction
operators, respectively. 
We denote by
$X$ and $Y$
the coupling constants of
a pair of additional interaction operators $\mathcal{O}^{X}$ and
$\mathcal{O}^{Y}$ that carry the same naive engineering dimension as
$\bar{\rho}^{2}+ \rho^{2}$ and $\bar{\rho} \rho$. Although these
additional operators are required to close the one-loop RG, we do not
define them explicitly 
at this point
(but see Sec.~\ref{IntOpsAnomDim}),
because they will turn out to be strongly
irrelevant in the disordered phase. To simplify notation, we define the
vector of interaction couplings
 \begin{equation}\label{IntVector}
	\vec{U} \equiv \{U,\,W,\,X,\,Y\}.
 \end{equation}

The expressions for the Schwinger-Keldysh path integral $Z$
[Eq.~(\ref{ZFull})] and the action $S_{D} + S_{I}$
[Eqs.~(\ref{SNonInt}) and (\ref{SInt})] constitute the completed
formulation of the 
full, interacting and disordered
fermion problem with chiral symmetry.
Upon averaging over realizations of the disorder potentials $m^{i}$ and
$A_{\mu}$, we find that the theory requires the specification of
a total of
seven
coupling constants: the random mass and random vector potential
variances $g$ and $g_{A}$, the four interaction strengths $\vec{U}$,
and finally the dynamic scale factor $h$. In the next section we
perform a perturbative one-loop RG analysis on this model. The
principal new results are the scaling behaviors of the interaction
couplings $\vec{U}$ as we look at lower and lower energies, or,
equivalently, larger and larger values of the flowing coupling $g_{A}$.
The reader interested only in our results may skip directly to Sec.\
\ref{Results}, where the scaling behaviors of the various couplings are
simply stated and analyzed.

\section{\label{1loop}ONE-LOOP CALCULATION}

In this section we perform a one-loop renormalization group analysis on
the interacting and disordered chiral fermion model defined in
Sec.~\ref{Setup}. We begin by re-writing the Grassmann Dirac field
appearing in the Keldysh path integral [Eq.~(\ref{ZFull})] in terms of
``left(L)-'' and ``right(R)-moving''
components, inspired by a similar decomposition made for the
non-interacting system in Ref.\ \onlinecite{GLL}. (This
L/R
decomposition provides a language similar to that used in Ref.\
\onlinecite{MRF} to investigate the 
freezing
transition in the dynamic
critical exponent and the low-energy density of states.) We then
perform a hard cutoff field theory renormalization group analysis to
one loop. The 
L/R
formulation allows for a particularly simple and
transparent treatment using diagrams. With a minimal effort we recover,
as a check, the one-loop RG results for the disorder-only
model\cite{HWK,BERNARD,GLL} [Eqs.\ (\ref{BetaFunchAnom},
\ref{BetaFuncgAnom}, \ref{BetaFuncgAAnom})]. Finally, we calculate
the anomalous dimensions of all allowed four-fermion interaction
operators in the presence of the disorder [Eqs.\ (\ref{BetaFuncWAnom},
\ref{BetaFuncXAnom}, \ref{BetaFuncYAnom}, \ref{BetaFuncUAnom})].

\subsection{Left(L)/Right(R) decomposition}\label{ChiralDecomp}

Consider the Keldysh action for the non-interacting disordered chiral
system, given in Eq.~(\ref{SNonInt}). We make the basis change
[$a=1,2$ is the Keldysh-index; 
Eqs.(\ref{SNonInt},\ref{SInt})]
\begin{equation}\label{UXform}
	\psi_{a} \rightarrow U \, \psi_{a},\, \bar{\psi}_{a} \rightarrow \bar{\psi}_{a} \, U^{\dagger}, 
 \end{equation} where the 
unitary
matrix $U$ is
\begin{equation}\label{U}
	U \equiv U_{p} \, \textstyle{\frac{1}{\sqrt{2}}} \left(1 + i \sigma^{2} \tau^{3} \right) \, \textstyle{\frac{1}{\sqrt{2}}} \left(1 - i \sigma^{2}\right).
\end{equation}
\newline
Here $U_{p} \equiv (1/2) \left( 1 + {\vec \sigma}\cdot {\vec \tau} \right)$
is the permutation operator, exchanging
$\sigma^{\alpha} \leftrightarrow \tau^{\alpha}$ upon conjugation.
This basis change induces a similarity transformation on the matrices 
defined in Eq.~(\ref{MatrixDef}); the transformed versions are 
$\bm{\alpha} =~(\alpha_{x},\alpha_{y}) =~(\sigma^{1} \tau^{1},\sigma^{2} \tau^{1})$ and 
$\{\beta^{0},\beta^{1},\beta^{2}\} =~\{\sigma^{3} \tau^{3} , \sigma^{3} \tau^{1} , \tau^{2} \}$.

Following Ref.\ \onlinecite{GLL}, we define the 
L/R
decomposition of
the Dirac
spinors:
 \begin{eqnarray}\label{ChiralSpinor}
	\psi_{a} & \equiv & \left[\begin{array}{c}  
					\bar{L}_{2 a} \\
					\bar{R}_{2 a} \\
					L_{1 a} \\
					R_{1 a} 
			\end{array}	
			\right]
 \end{eqnarray}
and
\begin{equation}\label{ChiralSpinorBar}
	\bar{\psi}_{a} \equiv s_{a} \left[\bar{R}_{1 a} \, \bar{L}_{1 a} \, R_{2 a} \, L_{2 a}  \right],
\end{equation}
where the first index $i \in \{1,2\}$ on the components $R_{i a}$ and $L_{i a}$ is a 
species index.
Thus the fields $R_{a}$ (``right-movers'') and $L_{a}$ (``left-movers'') are two-component 
objects. 
In terms of the latter, the non-interacting action given by Eq.~(\ref{SNonInt}) may now be written 
very simply as
\begin{widetext}
\begin{eqnarray}
S_{D} = &
i \, \int dt \, d^{2}r 
\left[
h \, \left (\bar{R}_{a} \mu^{1} \partial_{t} \bar{L}^{T}_{a}\right) 
+ h \,\left(L^{T}_{a} \mu^{1} \partial_{t} R_{a}\right) 
+ \bar{R}_{a} (\partial - A)R_{a} 
+ \bar{L}_{a}(\bar{\partial} - \bar{A})L_{a} \right. \nonumber \\
& \left. + m \, \bar{R}_{a} L_{a} 
+ \bar{m} \, \bar{L}_{a} R_{a} 
 \right]. \label{SNonIntChiral}
\end{eqnarray}
\end{widetext}
where the Pauli matrix $\mu^{1}$ is a matrix in the 
species space of the Fermion fields 
[which is the first index $i$ of Eqs.~(\ref{ChiralSpinor}) and 
(\ref{ChiralSpinorBar})].

In equation (\ref{SNonIntChiral}) we have suppressed the $T$ 
and $\bar{T}$-ordering pole prescriptions 
[Eq.(\ref{SNonInt})], and we have introduced the following 
complex notations: 
(i) $\partial \equiv \partial_{x} - i \partial_{y}$ and 
$\bar{\partial} \equiv \partial_{x} + i \partial_{y}$,
(ii) $A \equiv A_{y} + i A_{x}$ and 
$\bar{A} \equiv A_{y} - i A_{x}$ 
and (iii) $m \equiv m^{2} + i m^{1}$ and 
$\bar{m} \equiv m^{2} - i m^{1}$.

Turning to the interactions, we seek the set of all independent
four-fermion operators local in space and time with no derivatives; as
discussed in subsection \ref{Interactions}, these operators form a set
of local perturbations least irrelevant at the clean Dirac fixed point.
Any allowed interaction operator must be invariant under discrete
chiral, time-reversal, and 
spatial
reflection operations, and it must also
transform as a singlet under continuous U(1)
``electrical  charge'',
rotation,
and honeycomb lattice translation transformations.\footnote{The
continuum versions of honeycomb lattice rotations and translations
involve discrete ``rotations'' in the 4-component
(sublattice)$\times$(Fermi node) space.  
These rotations employ non-trivial discrete angles, so that we can generalize them to
continuous U(1) transformations in the low-energy theory.}
We summarize the action of these symmetry transformations in the above L/R
basis of low-energy fields in Table \ref{ChiralSymTable}. Consistent with
these requirements, we find the four independent local four-fermion operators
\begin{subequations}\label{IntOps}
\begin{eqnarray}
	\mathcal{O}^{A}_{a} & \equiv & 
(\bar{R}_{a} \frac{\mu^{1}}{2} R_{a}) \, 
(\bar{L}_{a} \frac{\mu^{1}}{2} L_{a}) 
+ 
(\bar{R}_{a} \frac{\mu^{2}}{2} R_{a}) \, 
(\bar{L}_{a} \frac{\mu^{2}}{2} L_{a}),\qquad \nonumber\\
 & & \\
	\mathcal{O}^{B}_{a} & \equiv & 
(\bar{R}_{a} \frac{\mu^{3}}{2} R_{a}) \, 
(\bar{L}_{a} \frac{\mu^{3}}{2} L_{a}), \\
	\mathcal{O}^{C}_{a} & \equiv & 
-\frac{(\bar{R}_{a} R_{a})}{2} \, \frac{(\bar{L}_{a} L_{a})}{2},
\end{eqnarray}
\end{subequations}
and
\begin{equation}\label{RhoBarRhoBar}
	\mathcal{O}^{U}_{a} \equiv \bar{\rho}_{a}^{2} + \rho_{a}^{2},
\end{equation}
where
\begin{subequations}\label{ChiralRhoandRhoBar}
\begin{eqnarray}
	\bar{\rho}_{a} & = & i 
(\bar{L}_{a} \mu^{2} \bar{R}_{a}^{T}), \\
	\rho_{a} & = & 
-i (R_{a}^{T} \mu^{2} L_{a}).
\end{eqnarray}
 \end{subequations} Equation (\ref{ChiralRhoandRhoBar}) gives the
L/R
language expressions for the long-wavelength sublattice density
operators defined by Eq.~(\ref{LocalDensity}). The operator
$\mathcal{O}^{U}_{a}$ given by Eq.~(\ref{RhoBarRhoBar}) is one of the
two four-fermion interaction terms that will turn out to be more
relevant in the disordered critical phase than at the clean Dirac fixed
point; the other, $\bar{\rho}_{a} \rho_{a}$, can be written in terms of
the three operators defined in Eq.~(\ref{IntOps}):
 \begin{equation}\label{RhoBarRhoChiralDecomp}
	\bar{\rho}_{a} \rho_{a} = -2 \left(\mathcal{O}^{A}_{a} + \mathcal{O}^{B}_{a} + \mathcal{O}^{C}_{a}\right).
 \end{equation}

Note that the three operators in Eq.~({\ref{IntOps}}), as well as the
whole structure of the non-interacting theory
[Eq.~(\ref{SNonIntChiral})] \emph{except} the ``energy'' terms (the
terms involving time derivatives) are invariant under 
a U(1) transformation, which we refer to as a ``sublattice'' U(1),
\begin{equation}\label{SubLatU1}
	R_{a} \rightarrow e^{i \theta} R_{a},\, L_{a} \rightarrow e^{i \theta} L_{a},\, \bar{R}_{a} \rightarrow \bar{R}_{a} e^{-i \theta},\, \bar{L}_{a} \rightarrow \bar{L}_{a} e^{-i \theta}.
 \end{equation}
(distinct from the ``electrical charge'' U(1) defined in Table I).
This U(1) transformation is \emph{not} a symmetry, because the energy
terms in the non-interacting action carry non-zero (i.e.~unit) charge.
In fact, under this $U(1)$,
the two energy terms $i (\bar{R}_{a} \mu^{1} \partial_{t}
\bar{L}_{a}^{T})$ and $i (L_{a}^{T} \mu^{1} \partial_{t} R_{a})$ carry
equal and opposite unit charges, as do the local sublattice density
operators $\bar{\rho}_{a}$ and $\rho_{a}$, while the four-fermion
interaction operator $\mathcal{O}_{a}^{U}$ [Eq.~(\ref{RhoBarRhoBar})]
is a sum of positive and negative doubly-charged same-sublattice
components. 

\begin{figure}
\includegraphics[width=0.4\textwidth]{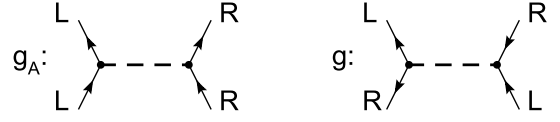}
\caption{Feynman rules for the diagrams in Fig.~\ref{FigSelfNRG}.\label{FigRulesA}}
\end{figure}

In the non-interacting disordered fermion problem, moments of
the local density of states (LDOS) operator are known to exhibit
negative scaling dimensions in the disordered,
critically delocalized phase.\cite{GLL,MRF} 
[The LDOS operator appears in the (2+1)D Keldysh
formulation as the Dirac fermion bilinear
$\bar{\psi}_{a}(\omega,\bm{r}) \psi_{a}(\omega,\bm{r})$, which is local
in position
space and frequency but not in time.] More precisely, it is the
\emph{components} of these LDOS moments with maximal charge under the
sublattice U(1) transformation defined by Eq.~(\ref{SubLatU1}) that
carry the maximally negative scaling dimensions, 
i.e.\ these components are the most relevant perturbations to the 
zero-energy nearly conformal field theory. The functional RG treatment 
employed in Ref.~\onlinecite{MRF} focuses just upon these most relevant 
components of the LDOS moments in order to extract the dynamic critical 
exponent in the regime above the freezing transition.
In our interacting problem, we will find that the 
interaction operator $\mathcal{O}_{a}^{U}=\bar{\rho}_{a}^2 +
\rho_{a}^{2}$,
which is maximally-charged under the ``sublattice'' U(1),
is the most relevant of the investigated four-fermion interaction
terms. Note that unlike the LDOS operator or the energy terms in the
non-interacting action (\ref{SNonIntChiral}), the sum of sublattice
density operators $\bar{\rho}_{a} + \rho_{a}$ \emph{cannot} receive any
corrections to scaling: this local spacetime operator is a component of
the $(2+1)$-dimensional conserved ``electrical charge''
U(1) current carried by the electrically-charged Dirac fermions
[Table I].

Next we average the Keldysh functional $Z$ [Eq.~(\ref{ZFull})] 
over realizations of the random vector potential $A=A_x+iA_y$ and the 
complex random 
mass $m$, with $\overline{\bar{A}(\bm{x})\,A(\bm{x'})} =4 \pi g_{A} \, \delta(\bm{x} - \bm{x'})$ and $\overline{\bar{m}(\bm{x})\,m(\bm{x'})} = 4 \pi g \, \delta(\bm{x} - \bm{x'})$. The resulting 
non-interacting disorder-averaged action 
$\overline{S_{D}}$ may be written in its final form as a sum of two terms: $\overline{S_{D}} \equiv S_{D}^{\,0} + \overline{\delta S_{D}}$, where
\begin{equation}\label{SNonInt0Chiral}
S_{D}^{\,0} = i \, \int dt \, d^{2}r \left[\bar{R}_{a} \, L_{a}^{T} \right] 
			\left[\begin{array}{cc}
				\partial & h \, \mu^{1} \partial_{t} \\
				h \, \mu^{1} \partial_{t} & \bar{\partial}
			\end{array}\right]
			\left[\begin{array}{c}
				R_{a} \\
				\bar{L}_{a}^{T} 
			\end{array}\right]\qquad	
\end{equation}
and
{\setlength\arraycolsep{.5pt}
\begin{eqnarray}\label{SNonIntDChiral}
	\overline{\delta S_{D}} = -4 \pi i \int dt \, dt' \, d^{2}r & & 
\left[ g \, (\bar{R}_{a} L_{a})(t) \, (\bar{L}_{b} R_{b})(t') \right. 
\nonumber \\
	& & \left.  +  g_{A} \, (\bar{R}_{a} R_{a})(t) \, 
(\bar{L}_{b} L_{b})(t')\right].\quad
\end{eqnarray}}The average over disorder has coupled together fermion 
bilinears with different Keldysh indices at arbitrarily distant times.

\begin{figure}
\includegraphics[width=0.4\textwidth]{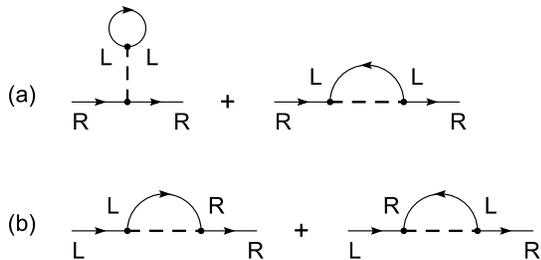}
\caption{Diagrams renormalizing the self-energy.\label{FigSelfNRG}}
\end{figure}

\subsection{Perturbative analysis}\label{Diagrams}

We calculate one-loop corrections to the propagators in 
(\ref{SNonInt0Chiral}) and to the disorder vertices in 
(\ref{SNonIntDChiral}), and we compute the anomalous dimensions 
of the interaction operators given by Eqs.~(\ref{IntOps}) and 
(\ref{RhoBarRhoBar}). The dependence of these corrections on a 
hard momentum cutoff $\Lambda$ will be
used\cite{DJA} to derive the renormalization group equation summarized
in Sect. IV.

\subsubsection{Self-energy}\label{Self-energy}

In order to calculate the self-energy, it is useful to define the 
fields $\chi_{a}$ and $\bar{\chi}_{a}$ via
[Keldysh index $a \in \{1,2\}$]
\begin{equation}\label{Chi}
	\chi_{a} \equiv \left[\begin{array}{c}
					R_{a} \\
					\bar{L}_{a}^{T}
				\end{array}\right]
\end{equation}
and
\begin{equation}\label{ChiBar}
	\bar{\chi}_{a} \equiv \left[\bar{R}_{a} \, L_{a}^{T} \right].
\end{equation}
We can think of $\chi_{a} \equiv \left\{\chi_{\gamma a}\right\}_{\gamma}$ as a 
2-component 
R/L-valued object; the index $\gamma \in \{R,L\}$ 
addresses the two components of the definition in 
Eq.~(\ref{Chi}); each field $R_{a}$ or $\bar{L}_{a}$ carries 
in addition a 2-component species index $i$, defined in
Eq. (\ref{ChiralSpinor}). ($\chi_{\gamma i a}$ when all indices are
displayed.)

The propagator for the field  $\chi_{a}$ follows from Eq.~(\ref{SNonInt0Chiral}), 
and is given by 
\begin{equation}\label{ChiProp}
	\left\langle \chi_{a}(\omega,\bm{k}) \bar{\chi}_{a}(\omega,\bm{k}) \right\rangle  = \frac{-i}{\bar{k} k - h^{2}(\omega^{2} + i \eta s_{a})} 
												\begin{bmatrix}
													\bar{k} & h \omega \mu^{1} \\
													h \omega \mu^{1} & k
												\end{bmatrix}\!\!,
\end{equation}
where $k \equiv k_{x} - i k_{y}$ and $\bar{k} \equiv k_{x} + i
k_{y}$. In Eq.~(\ref{ChiProp}) we have restored the correct $T$ and
$\bar{T}$-ordering pole prescriptions in the two Keldysh species. We
can also re-write the disorder-averaged action $\overline{\delta
S_{D}}$ given by Eq.~(\ref{SNonIntDChiral}) in terms of the
$\chi_{\gamma a}$ fields
[``T'' denoting the transpose, and $\gamma \in \{ L,R\}$]:
{\setlength\arraycolsep{.5pt}
\begin{eqnarray}\label{SNonIntDChiralChi}
	\overline{\delta S_{D}} = -4 \pi i \int & & dt \, dt' \, d^{2}r 
\left[ g \, (\bar{\chi}_{R a} \, \bar{\chi}_{L a}^{T})(t) \, 
(\chi_{L b}^{T} \, \chi_{R b})(t') \right. \nonumber \\
	& & \left.  -  g_{A} \, (\bar{\chi}_{R a} \, \chi_{R a})(t) \, 
(\bar{\chi}_{L b} \, \chi_{L b})(t')\right].\quad
\end{eqnarray}} 
We depict the disorder vertices $g$ and $g_{A}$ with dashed lines, shown in Fig.~\ref{FigRulesA}. 

Figure \ref{FigSelfNRG} shows the one-loop disorder contributions to the self-energy. In these diagrams the propagator $\left\langle \chi_{\gamma} \bar{\chi}_{\gamma'} \right\rangle$ is represented as a directed solid line labeled with the endpoint indices $(\gamma, \, \gamma') \in \{R,L\}$. (Keldysh indices are suppressed.)  We furnish the external $\chi$ field legs with the labels ``R'' and ``L'' to indicate which propagator is being corrected.

\begin{figure}
\includegraphics[width=0.4\textwidth]{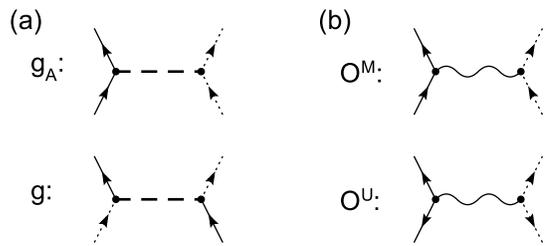}
\caption{Feynman rules for the diagrams in Figs.~\ref{FiggRenorm}---\ref{FigRhoBarRhoBarRenorm}. 
The interaction vertex labeled $\mathcal{O}^{M}$ denotes the sublattice U(1) charge-neutral operators defined by Eq.~(\ref{IntOps}), while the vertex labeled $\mathcal{O}^{U}$ depicts the $\bar{\rho}^{2}$ component of the sublattice-charged interaction operator $\bar{\rho}^{2} + {\rho}^{2}$. \label{FigRulesB}}
\end{figure}

The two diagrams appearing in Fig.~\ref{FigSelfNRG}(a) correspond to the one-loop self-energy corrections to $\left\langle \chi_{R} \bar{\chi}_{R} \right\rangle$. Both of these diagrams vanish: the first diagram contributes nothing because it contains a free Keldysh index summation,\cite{KELDYSH} 
while the second diagram vanishes because the momentum integral of the 
purely left-moving propagator 
$\left\langle \chi_{L} \bar{\chi}_{L} \right\rangle$ 
gives zero. Thus we find that there is no self-energy correction to the
 ``R-R'' sector of the $\chi$ field propagator
to one-loop order.

By contrast, the diagrams contributing to the self-energy correction 
to $\left\langle \chi_{R} \bar{\chi}_{L} \right\rangle$ shown in 
Fig.~\ref{FigSelfNRG}(b) lead to a non-zero renormalization of the 
dynamic scale factor $h$. The irreducible two-point vertex function 
in this channel to one loop is
[cf.\  Eq.~(\ref{ChiProp})]
\begin{equation}\label{Gammah}
	\Gamma_{R L}(\omega,\bm{k}) = i h \omega \mu^{1}[1 + 2(g + g_{A})\ln\Lambda], 
\end{equation}
where $\Lambda$ is an ultraviolet momentum cutoff and we 
did not write out ultraviolet-finite corrections. Differentiating Eq.~(\ref{Gammah}) 
with respect to $\ln\Lambda$ and setting the result equal to zero 
leads to the expression\cite{DJA}
\begin{equation}\label{BetaFunchAnom}
	\left( \frac{d \ln h}{d l} \right)_{an} =  2(g + g_{A}),
\end{equation} 
where $l \equiv  -\ln\Lambda$ is the logarithm of the RG length scale. 
The subscript ``$an$''
appearing on the left hand side of 
Eq.~(\ref{BetaFunchAnom}) stands for ``anomalous,'' as the above 
calculation has given us only the anomalous part of the beta function; 
in order to obtain the full one-loop flow equation for $h$, we must add 
the appropriate 
``engineering dimension''
to Eq.~(\ref{BetaFunchAnom}). 
This will be done in Sec.~\ref{Results}, below.

Although we have considered the propagator renormalization due only to the 
disorder, it can be easily seen that none of the local spacetime interaction 
operators defined in Eqs.~(\ref{IntOps}) and (\ref{RhoBarRhoBar}) can generate 
a non-zero one-loop contribution to the Dirac field
self-energy; such a contribution would necessarily correspond to either the 
generation of a mass or a chemical potential shift, both forbidden by symmetry.

\subsubsection{Disorder self-renormalization}\label{DisorderSelfRenorm}

Next we turn to the self-renormalization of the disorder couplings 
$g$ and $g_{A}$. We will only need the diagonal components of the 
propagator given by Eq.~(\ref{ChiProp}), so we
revert back to the $\{R,L,\bar{R},\bar{L}\}$ 
component language [the species index $i=1,2$, Eq. 
(\ref{ChiralSpinor}, \ref{ChiralSpinorBar}), will be suppressed].
The propagators we will need are
\begin{equation}\label{RRProp}
	\left\langle R_{a}(\omega,\bm{k}) \bar{R}_{a}(\omega,\bm{k}) \right\rangle  = \frac{-i \bar{k}}{\bar{k} k - h^{2}(\omega^{2} + i \eta s_{a})} 
\end{equation}
and
\begin{equation}\label{LLProp}
	\left\langle L_{a}(\omega,\bm{k}) \bar{L}_{a}(\omega,\bm{k}) \right\rangle  = \frac{-i k}{\bar{k} k - h^{2}(\omega^{2} + i \eta s_{a})}. 
\end{equation}
Note that both propagators are odd under the exchange $k \rightarrow -k$.

Now we will use directed solid and dotted lines to depict the 
$\left\langle R \bar{R} \right\rangle$ and $\left\langle L \bar{L} \right\rangle$ 
propagators, respectively. We will continue to use dashed lines to 
indicate the disorder vertices $g$ and $g_{A}$, as shown in Fig.~\ref{FigRulesB}(a). 

\begin{figure}
\includegraphics[width=0.4\textwidth]{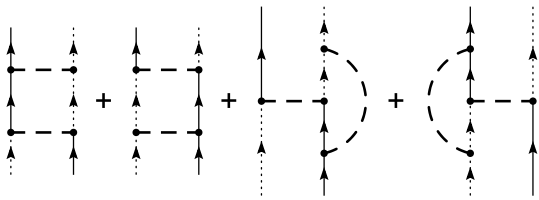}
\caption{Diagrams renormalizing $g$.\label{FiggRenorm}}
\end{figure}

Figure \ref{FiggRenorm} shows the corrections to the $g$ vertex
in Eq. (\ref{SNonIntDChiral}).
The $\ln\Lambda$ contributions due to the first two diagrams exactly cancel those of the second two, leading to the anomalous
part of the
one-loop beta function
\begin{equation}\label{BetaFuncgAnom}
	\left(\frac{d g}{d l}\right)_{an} = 0.
\end{equation}
The cancelation is easily understood if one notes that all four diagrams in Fig.~\ref{FiggRenorm} contain a closed momentum loop integration involving one of each diagonal propagator given by Eqs.~(\ref{RRProp}) and (\ref{LLProp}). All four diagrams contribute the same absolute ultraviolet-divergent amplitude to the sum. Note that the loop momentum necessarily flows against one of the two directed chiral propagator arrows in the first pair of diagrams in Fig.~\ref{FiggRenorm}. In the second pair of diagrams, the loop momentum can be taken to flow in the same direction as both 
L- and R-propagator arrows. This simple structural difference produces a relative minus sign, forbidding the coupling $g_{A}$ from feeding into $g$. The same mechanism will lead to cancelations in the autorenormalization of $g_{A}$ as well as in the computation of the anomalous dimensions of some of the interaction operators. 

In Figure \ref{FiggARenorm}, we depict the one-loop disorder-disorder diagrams that renormalize the coupling $g_{A}$. The first two diagrams, each proportional to $g_{A}^{2}$, exactly cancel via the mechanism described above. The third diagram is non-zero, and leads to the irreducible four-point vertex function
\begin{equation}\label{GammagA}
	\Gamma_{g_{A}} = g_{A} + 2 g^{2} \ln\Lambda.
\end{equation}
Differentiation with respect to $l = -\ln\Lambda$ produces the anomalous
part of the
one-loop beta function
\begin{equation}\label{BetaFuncgAAnom}
	\left(\frac{d g_{A}}{d l}\right)_{an} =  2g^{2}.
\end{equation}

\begin{figure}
\includegraphics[width=0.4\textwidth]{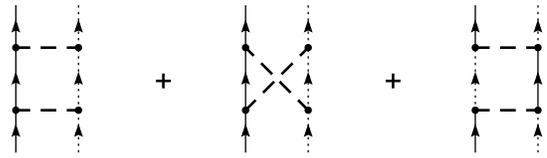}
\caption{Diagrams renormalizing $g_{A}$.\label{FiggARenorm}}
\end{figure}

\subsubsection{Interaction operator anomalous dimensions}\label{IntOpsAnomDim}

Finally, we turn to the one-loop calculation of the anomalous dimensions of the 
four interaction operators defined in Eqs.~(\ref{IntOps}) 
and (\ref{RhoBarRhoBar}) in the disordered (critical)
phase characterized by finite $g$ and $g_{A}$.
The results are listed
in Eqs.~(\ref{BetaFuncWAnom}, \ref{BetaFuncXAnom},
\ref{BetaFuncYAnom}, \ref{BetaFuncUAnom})
and 
Eqs.~(\ref{FlowEqU}, \ref{FlowEqW}, \ref{FlowEqX}, \ref{FlowEqY}).

Consider first the three ``sublattice-U(1)'' charge-neutral 
interaction operators defined in Eq.~(\ref{IntOps}). We will 
use a wavy line to represent all three interaction operators 
simultaneously, as depicted by the vertex labeled 
$\mathcal{O}^{M}$ in Fig.~\ref{FigRulesB}(b); these operators differ 
only in their species index structure.  Figure \ref{FigIntOpsRenorm}
shows the one-loop renormalization of these interaction operators due
to the disorder vertices $g$ and $g_{A}$. The labels $i,j$ and $k,l$
appearing on the external legs denote the species components of the outgoing and incoming right and left-moving 
particles, respectively. The first four diagrams describe the $g_{A}$-dressing of the interactions; these diagrams cancel up to ultraviolet-finite contributions. The final two diagrams in Fig.~\ref{FigIntOpsRenorm} involve the $g$ disorder vertex and add constructively. Let us denote the \emph{zeroth} order irreducible four-point vertex function for the
 interaction operator $\mathcal{O}^{M}$ via
\begin{equation}\label{GammaIntOps0}
	\Gamma^{\,0}_{M \, ij;kl} \equiv \sum_{\lambda} M_{ij}^{\lambda} M_{kl}^{\lambda}. 
\end{equation}
In this equation, the vertex matrices $M^{\lambda}$ depend upon the operator in question, e.g.\ the operator $\mathcal{O}^{A}$ [Eq.~(\ref{IntOps}a)] has two such vertex matrices, $M^{1} = \mu^{1}/2$ and $M^{2} = \mu^{2}/2$. Incorporating the correction due to $g$, we find the one-loop irreducible four-point vertex function for the operator $\mathcal{O^{M}}$ 
\begin{equation}\label{GammaIntOpsM}
	\Gamma_{M \, ij;kl} = \sum_{\lambda} M_{ij}^{\lambda} M_{kl}^{\lambda}  - 4 g \ln\Lambda  \sum_{\lambda} M_{il}^{\lambda} M_{kj}^{\lambda}. 
\end{equation}
We have such a vertex function for each 
of the three operators defined in Eq.~(\ref{IntOps}).

\begin{figure}
\includegraphics[width=0.4\textwidth]{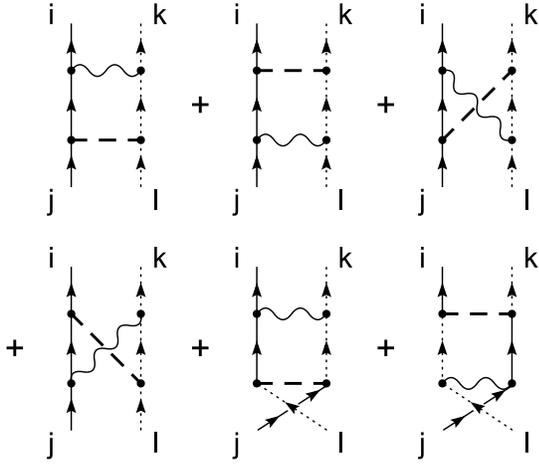}
\caption{Diagrams renormalizing the interaction operators defined by Eq.~(\ref{IntOps}).\label{FigIntOpsRenorm}}
\end{figure}

We rewrite Eq.~(\ref{GammaIntOpsM}) using the Pauli matrix identities
\begin{subequations}
\begin{equation}
	\mu^{1}_{il}\mu^{1}_{kj} + \mu^{2}_{il}\mu^{2}_{kj}  =  \delta_{ij}\delta_{kl} - \mu^{3}_{ij}\mu^{3}_{kl},
\end{equation}
\begin{equation}
	\mu^{3}_{il}\mu^{3}_{kj}  =  -\textstyle{\frac{1}{2}}\left(\mu^{1}_{ij}\mu^{1}_{kl} + \mu^{2}_{ij}\mu^{2}_{kl} \right) + 
	\textstyle{\frac{1}{2}}\left( \delta_{ij}\delta_{kl} + \mu^{3}_{ij}\mu^{3}_{kl} \right),
\end{equation}	
\text{and}
\begin{equation}
	\delta_{il}\delta_{kj} =  \textstyle(\frac{1}{2}) \left( \delta_{ij}\delta_{kl} + \mu^{\alpha}_{ij}\mu^{\alpha}_{kl} \right),  
\end{equation}
\end{subequations}
where in this last identity $\alpha$ is summed over $\{1,2,3\}$. 
Using these identities with Eq.~(\ref{GammaIntOpsM}) shows us 
that the interaction operators mix 
under disorder-renormalization,
leading to the matrix vertex equation
\begin{equation}\label{IntOpsMatrixEquation}
	\left[\begin{array}{c}
		\Gamma_{A} \\
		\Gamma_{B} \\
		\Gamma_{C} 
	\end{array}\right]
	=
	\left[\begin{array}{ccc}
		1 & 2c & 2c \\
		c & 1-c & c \\
		c & c & 1-c
	\end{array}\right]
	\left[\begin{array}{c}
		\Gamma^{\,0}_{A} \\
		\Gamma^{\,0}_{B} \\
		\Gamma^{\,0}_{C}
	\end{array}\right],
\end{equation}
where the constant $c \equiv 2 g \ln\Lambda$.

The eigenvalues of the matrix in Eq.~(\ref{IntOpsMatrixEquation}) determine the anomalous dimensions of the RG eigenoperators, which are linear combinations of the interaction operators defined in Eq.~(\ref{IntOps}). These eigenvalues are $1 + 4 g \ln\Lambda$ and $1 - 4 g \ln\Lambda$; the latter is doubly degenerate. The eigenoperator corresponding to the eigenvalue $1 + 4 g \ln\Lambda$ is
\begin{equation}\label{RhoBarRho}
	\mathcal{O}^{A} + \mathcal{O}^{B} + \mathcal{O}^{C} \propto \bar{\rho}\rho. 
\end{equation}
If we add this interaction operator to the action of the theory 
with a coupling constant $W$, as in Eq.~(\ref{SInt}), then we find the 
anomalous part of the beta function for $W$
\begin{equation}
\label{BetaFuncWAnom}
	\left(\frac{d W}{d l}\right)_{an} = 4 g W, 
\end{equation}
making the interaction coupling $W$ more relevant in the (critical) 
disordered phase than at the clean Dirac fixed point. The other two eigenoperators share the RG eigenvalue $1 - 4 g \ln\Lambda$; 
if we assign these operators the 
coupling constants
$X$ and $Y$, then we have
\begin{equation}
\label{BetaFuncXAnom}
	\left(\frac{d X}{d l}\right)_{an} = -4 g X, 
\end{equation}
and
\begin{equation}
\label{BetaFuncYAnom}
	\left(\frac{d Y}{d l}\right)_{an} = -4 g Y, 
\end{equation}
so that the couplings $X$ and $Y$ turn out to be more irrelevant in the presence of disorder.

\begin{figure}
\includegraphics[width=0.4\textwidth]{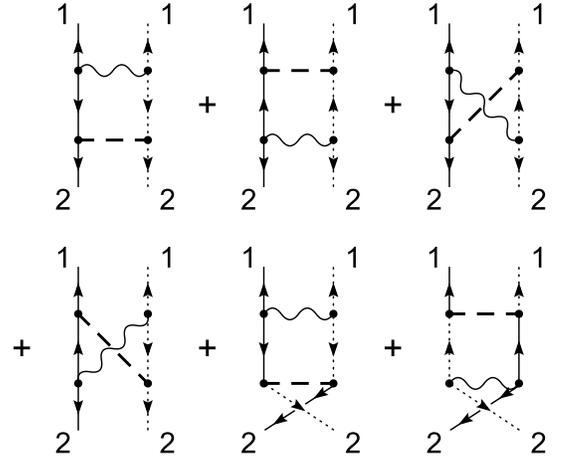}
\caption{Diagrams renormalizing the interaction operator $\bar{\rho}^{2}$.\label{FigRhoBarRhoBarRenorm}}
\end{figure}

Finally, we turn to the same-sublattice interaction operator 
$\mathcal{O}^{U}$ given by Eq.~(\ref{RhoBarRhoBar}). 
We will use diagrams to compute the renormalization of the 
$\bar{\rho}^{2}$ half of this operator; the other half $\rho^{2}$ 
must renormalize the same way. 
Using Eq.~(\ref{ChiralRhoandRhoBar}), 
we write 
$\bar{\rho}^{2} = - \left(\bar{L}_a \mu^{2} \bar{R}_a^{T}\right)^{2} 
\propto \bar{L}_{1a} \bar{L}_{2a} \bar{R}_{1a} \bar{R}_{2a}$. 
We will depict the 
interaction $\bar{\rho}^{2}$ as a wavy line again joining right and 
left-mover lines, but now all such lines are directed outwards 
from the vertex---this half of the interaction operator is a 
source of
``sublattice'' U(1) current
[see Fig.~\ref{FigRulesB}(b)]. 

The diagrams describing the disorder-dressing of the $\bar{\rho}^{2}$ 
operator are shown in Fig.~{\ref{FigRhoBarRhoBarRenorm}}. 
Numerical labels on the external legs again refer to 
the species index $i,j=\{1,2\}$.
The first four diagrams involve dressing by the 
$g_{A}$ vertex. The key point is that these four diagrams add 
constructively for this sublattice-charged interaction: $g_{A}$ feeds 
into the operator $\bar{\rho}^{2}$. The cancelation that occured for 
the sublattice charge-neutral operators given by Eq.~(\ref{IntOps}) does not occur here, because all four diagrams proportional to $g_A$ involve a loop momentum that necessarily flows against one of the two directed 
propagator arrows in that loop. The final two diagrams in Fig.~{\ref{FigRhoBarRhoBarRenorm}} depict the $g$-renormalization of the interaction. Summing these contributions we obtain the irreducible four-point vertex function for the interaction operator $\mathcal{O}^{U} = \bar{\rho}^{2} + \rho^{2}$ to one loop
\begin{equation}\label{GammaRhoBarRhoBar}
	\Gamma_{U} = 1 + 4 \left(2 g_{A} - g \right)\ln\Lambda.
\end{equation}  
Adding $\mathcal{O}^{U}$ to the action with coupling $U$ as 
in Eq.~(\ref{SInt}), we find the anomalous
part of the
beta function for the 
coupling $U$ given by
\begin{equation}
\label{BetaFuncUAnom}
	\left(\frac{d U}{dl}\right)_{an}	
= 4 \left(2 g_{A} - g \right) U.
\end{equation}

\section{\label{Results}RESULTS AND DISCUSSION}

\subsection{Scaling equations}

In this section we summarize the scaling behaviors of the coupling constants for the model defined by Eqs.~(\ref{ZFull}), (\ref{SNonInt}), and (\ref{SInt}) in Sec.~\ref{Setup}. Seven coupling constants are needed to describe that model: these are the disorder variances $g$ and $g_{A}$, the interaction strengths $\vec{U}~=~\left\{U,\,W,\,X,\,Y \right\}$, and the dynamical scale factor $h$. We may extract the
full RG beta functions for
these couplings by combining the results of Sec.~\ref{1loop} with 
dimensional analysis,
yielding the ``engineering dimensions.''
Examine the action for the theory given by Eqs.~(\ref{SNonInt}) and 
(\ref{SInt}). We take the dependence of the energy ``$\omega$'' 
upon the RG scale factor ``$l$''
[the logarithm of the (spatial) length scale]
to be
\begin{equation}\label{OmegaFlowDef}
	\frac{d \ln \omega}{d l} \equiv z(l),
\end{equation}
In Eq.\ (\ref{OmegaFlowDef}), $z$ is the (possibly scale-dependent) ``dynamic critical exponent.'' We say that energy has the ``engineering dimension''
\begin{equation}\label{OmegaDim}
	\left[ \omega \right] = z
\end{equation} 
in inverse length units. We take the Fermi velocity $v_{F} = 1$ to be dimensionless; then, from Eq.~(\ref{SNonInt}), the Dirac field has the engineering dimension
\begin{equation}\label{PsiDim}
	\left[\psi(t,\bm{r})\right] = \frac{1 + z}{2},
\qquad
	\left[\psi(\omega,\bm{r})\right] = \frac{1 - z}{2},
\end{equation}
in two spatial dimensions. 
We then find that the disorder strengths
$g$ and $g_{A}$ are dimensionless, the interaction couplings $\vec{U}$ share the engineering dimension
\begin{equation}\label{IntDim}
	[\vec{U}] = -z,
\end{equation}
and the dynamic scale factor $h$ has the dimension
\begin{equation}\label{hDim}
	\left[h\right] = 1 - z. 
\end{equation} 

Combining the anomalous dimensions in Eqs.~(\ref{BetaFunchAnom}), (\ref{BetaFuncgAnom}), (\ref{BetaFuncgAAnom}), (\ref{BetaFuncWAnom}-\ref{BetaFuncYAnom}), and (\ref{BetaFuncUAnom}) from Sec.~\ref{1loop} with the above engineering dimensional analysis, we find the one-loop flow equations
\begin{align}
	\frac{d \ln h}{d l} & = (1 - z) + 2(g_{A} + g) \nonumber \\
				& \phantom{==} + \mathcal{O}(g^2, g g_{A}, g \vec{U}, g_{A} \vec{U}, \vec{U}^{2}), \label{FlowEqh} \\
	\frac{d g}{d l} & =  0 + \mathcal{O}(g \vec{U},
g^3
), \label{FlowEqg}\\
\intertext{and}
	\frac{d g_{A}}{d l} & = 2 g^{2} + \mathcal{O}(g_{A} \vec{U}, g^{3}) \label{FlowEqgA},
\end{align}
for the parameters that appear in the non-interacting sector 
of the theory, and 
\begin{subequations}
\label{FlowEqsInt}
\begin{align}
	\frac{d U}{d l} & 
= (-z + 8 g_{A} - 4 g) U + \mathcal{O}(\vec{U}^2, g^{2} U, g g_{A} U) 
\label{FlowEqU}\\
	\frac{d W}{d l} & 
= (-z + 4 g) W + \mathcal{O}(\vec{U}^2, g^{2} W) 
\label{FlowEqW}\\
	\frac{d X}{d l} & 
= (-z - 4 g) X + \mathcal{O}(\vec{U}^2, g^{2} X) 
\label{FlowEqX}\\
\intertext{
as well as
}
	\frac{d Y}{d l} & 
= (-z - 4 g) Y + \mathcal{O}(\vec{U}^2, g^{2} Y) 
\label{FlowEqY},
\end{align}
\end{subequations}
for the interactions.
\footnote{
The lack of higher order corrections in $g_{A}$ to Eqs.~(\ref{FlowEqh}-\ref{FlowEqsInt}) is discussed in detail below.}

In Eqs.~(\ref{FlowEqh}-\ref{FlowEqgA}) and (\ref{FlowEqsInt}), the RG function $z(l)$ is so far undetermined. We may choose $z$ any way we like; different choices correspond to different ways of rescaling energy in the (2+1)D field theory after performing the RG transformation. The natural choice 
\begin{equation}\label{FlowEqh2}
	\frac{d \ln h}{d l} \equiv 0,
\end{equation} 
eliminates the parameter $h$ from the theory.
Enforcing Eq.~(\ref{FlowEqh2}) order by order in perturbation theory uniquely determines $z$; to one loop, we find that
\begin{equation}\label{ZDef}
	z(l) = 1 + 2(g_{A} + g).
\end{equation} 

First, note (as a check) that we recover the known results\cite{HWK,BERNARD,GLL} (\ref{FlowEqg}), (\ref{FlowEqgA}), and (\ref{ZDef}) for the non-interacting sector of the theory: the random mass variance $g$ is purely marginal to one loop; by contrast, $g_{A}$ is driven to strong coupling by $g \neq 0$; the scale-dependent dynamic critical exponent $z$ is fed by both $g$ and $g_{A}$.

The scaling behaviors of the interaction couplings given in Eq.~(\ref{FlowEqsInt}) are new and will be analyzed below. We stress that the terms maintained in the flow equations (\ref{FlowEqg}-\ref{FlowEqsInt}) and (\ref{ZDef}) are only those required to determine the stability of the non-interacting 
phase.\footnote{The full one-loop RG, which we do not present in 
detail here, includes mixed interaction-disorder terms that renormalize $g$ 
[$\mathcal{O}(g \vec{U})$ corrections to Eq.~(\ref{FlowEqg})] 
and $g_{A}$ [$\mathcal{O}(g_{A} \vec{U})$ corrections to 
Eq.~(\ref{FlowEqgA})], and interaction self-renormalization 
effects [$\mathcal{O}(\vec{U}^{2})$ corrections to 
Eq.~(\ref{FlowEqsInt})].}

Now we address an immediate concern: we are interested in the disordered critical phase that occurs near zero energy as $g_{A}$ flows away from the clean Dirac fixed point (where $g = g_{A} = 0$) to strong coupling; this flow always occurs for $g \neq 0$. The reader 
should 
therefore question the value of the purely perturbative one-loop results that we have presented above; naively, these lowest order results cannot be trusted to tell us anything about the strong coupling large-$g_{A}$ regime. 

Nevertheless, we can safely extend our results to large values of the 
random vector potential variance $g_{A} \gg 1$; in order to do so, we 
rely upon several known exact results for the 
disorder-only model\cite{GLL,MRF}. In the absence of interactions, the 
exact flow equation for $g$ is strictly zero, while the exact flow 
equation for $g_{A}$ depends only upon $g$ and vanishes for $g=0$. 
The exact flow equation for the dynamical exponent $z$ \emph{below the 
freezing transition} is linear in $g_A$, with non-linear $g$-dependent coefficients. 
Equations (\ref{FlowEqgA}) and (\ref{ZDef}) give the correct behaviors 
to lowest order in $g$. In this paper, we imagine tuning $g \ll 1$; 
then our results in fact tell the whole story for the couplings $g$ and $g_{A}$, 
and for the dynamic critical exponent $z$. (Above the freezing transition, 
with $g_{A} > g_{A}^{\, c}$, we will need to be more careful---see below.)

Indeed, the 
scaling dimensions of the important 
four-fermion interaction operators $\bar{\rho}^{2} + \rho^{2}$ and $\bar{\rho} \rho$, 
given to one loop in Eqs.~(\ref{FlowEqU}) and (\ref{FlowEqW}), respectively (the terms linear in $\vec{U}$), 
must be \emph{exactly} linear in $g_{A}$, with higher-loop corrections only modifying their 
$g$-dependences. 
These scaling dimensions combine the engineering dimension ``$-z$''
in these equations
with the anomalous dimensions calculated in Sec.~\ref{1loop}. The anomalous dimensions 
characterize the disorder-dressing of the interaction operators at zero interaction coupling, 
and could alternatively be calculated in the $(2+0)$D (non-interacting) version of the theory. Such a calculation in the $(2+0)$D theory can be performed non-perturbatively in $g_{A}$\cite{GLL,MRF}; the anomalous dimension of a local operator in this $(2+0)$D theory is guaranteed to be linear in $g_{A}$ with a coefficient determined by the ``sublattice'' [Eq.~(\ref{SubLatU1})] and ``translation'' [see Table \ref{ChiralSymTable}] U(1) charges carried by that operator. Therefore our results in Eqs.~(\ref{FlowEqU}) and (\ref{FlowEqW}) are in fact valid to all orders in $g_{A}$. Eqs.~(\ref{FlowEqU}) and (\ref{FlowEqW}) incorporate, in addition, the lowest order corrections in $g$. 

In the following, we will analyze our results. We note from 
Eq.~(\ref{FlowEqsInt}) that the interaction couplings $U$ and $W$ 
are enhanced by the disorder, while $X$ and $Y$ are (weakly) 
suppressed relative to the clean Dirac fixed point. We therefore focus 
our attention upon the former pair of interactions. Integrating the flow 
equations (\ref{FlowEqU}) and (\ref{FlowEqW}), we find
using Eq.~(\ref{BetaFuncUAnom})
\begin{align}\label{UEvolve}
	U(l) & \sim 
\left(\frac{\omega(0)}{\omega(l)}\right) 
\exp\left[\int^{l}_{0} dl'
\left ( {\frac{dU}{dl'}}\right )_{an}
\right] \nonumber \\
		& \sim \left(\frac{\omega(0)}{\omega(l)}\right) \exp\left[2 (2 g \, l)^{2} + (8 g_A^{\,0} - 4 g) l \right] \\
\intertext{and using Eq.~(\ref{BetaFuncWAnom})} \label{WEvolve}
	W(l) & 
\sim \left(\frac{\omega(0)}{\omega(l)}\right) 
\exp\left[\int^{l}_{0} dl' 
\left ( {\frac{dW}{dl'}}\right)_{an}
\right] \nonumber \\
		& \sim \left(\frac{\omega(0)}{\omega(l)}\right) \exp\left[4 g \, l \right],
\end{align}
where we take $g_{A}^{\,0} \ll 1$ as the bare value of $g_{A}$. In order to derive Eqs.~(\ref{UEvolve}) and (\ref{WEvolve}), we have integrated the flow equations for $g$ and $g_{A}$ given by Eqs.~(\ref{FlowEqg}) and (\ref{FlowEqgA}).

To complete the story, we need the relationship between the 
energy $\omega(l)$ and the RG parameter $l$; we obtain this by 
integrating Eq.~(\ref{ZDef}). We solve the resulting equation for $l$ 
in terms of $\ln \omega$, where $\omega = \omega(0)$ is the bare energy 
scale. The result is
{\setlength\arraycolsep{.5pt}
\begin{eqnarray}\label{DynamicScaling}
	l \sim \frac{1}{(2g)^{2}} & & \left\{-1 - 2\left(g + g_{A}^{\,0}\right) \phantom{\sqrt{\left[g_{A}^{\,0}\right]^{2}}}\right. \nonumber \\ 			& & \left. + \sqrt{\left[1 + 2\left(g + g_{A}^{\,0}\right)\right]^{2} + 8 g^{2} \left|\ln \frac{\omega}{\Omega} \right|} \right\},\qquad
\end{eqnarray}}  
where $\Omega \equiv \omega(l) \geq \omega$ is an arbitrary finite reference energy. We will state all of our results in terms of the bare energy $\omega \rightarrow 0$, so it is useful to note that the coupling $g_A$ 
evolves according to
\begin{equation}\label{gAEvolve}
	g_{A}(l) \sim 2 g^{2} l + g_{A}^{\,0}, 
\end{equation}
and we use Eq.~(\ref{DynamicScaling}) to relate $l$ to $\omega$.

Now we examine the scaling behaviors of the interaction strengths given by Eqs.~(\ref{UEvolve}) and (\ref{WEvolve}) in the 
following three dynamical regimes, which we distinguish by the strength 
of the vector potential disorder $g_{A}$: 
(1)~near the clean Dirac fixed point ($g_{A}~\sim~g_{A}^{\,0}~\ll~1$), (2)~at intermediate disorder below 
the freezing transition ($g_{A} \rightarrow g_{A}^{\, c}$), and 
finally (3)~above the transition at asymptotically strong disorder ($g_{A} \gg g_{A}^{\, c}$). 
The freezing transition occurs\cite{MRF} at $g_{A} = 1 + \mathcal{O}(g)$ for small
$g$. The idea is to tune both $g$ and $g_{A}$ small at some 
arbitrary finite energy scale $\omega = \Omega$; this places us close 
to the clean Dirac fixed point, where we know the interactions 
characterized by $U$ and $W$ are both strongly irrelevant. We then examine 
successively lower and lower energies $\omega \rightarrow 0$; as we do so, 
the RG sends $g_{A}$ to larger and larger values, and we expect 
scaling behavior of the interaction operators to change as we 
approach the delocalized, critical phase at zero energy.

We now discuss in turn the regimes (1), (2), and (3) defined above:

\subsubsection{Near the clean Dirac fixed point: $g_{A} \ll 1$}\label{WeakDisorder}

We begin near the clean Dirac fixed point with $g^{2} \left|\ln (\omega / \Omega) \right| \ll 1$. We can then approximate the dynamic scaling relationship (\ref{DynamicScaling}) as
\begin{equation}\label{DynamicScalingWeak}
	l \sim \left| \ln \frac{\omega}{\Omega} \right| \left(1 - 2 g - 2 g_{A}^{\,0} \right),
\end{equation}
which corresponds to a dynamic critical exponent $z \sim 1 + 2 g + 2 g_{A}^{\,0}$, implying a vanishing density of states. Inserting Eq.~(\ref{DynamicScalingWeak}) into Eq.~(\ref{gAEvolve}) gives the consistency check $g_{A} \ll 1$. 

Using Eq.~(\ref{DynamicScalingWeak}), the scaling equations (\ref{UEvolve}) and (\ref{WEvolve}) for the couplings $U$ and $W$ become
\begin{equation}\label{UEvolveWeak}
	U \sim \left(\frac{\omega}{\Omega}\right)^{1 + 4 g - 8 g_{A}^{\,0}},
\end{equation}
and
\begin{equation}\label{WEvolveWeak}
	W \sim \left(\frac{\omega}{\Omega}\right)^{1 - 4 g}.
\end{equation}
As expected, near the clean Dirac fixed point both interactions are strongly irrelevant as we look at decreasing energy scales.

\subsubsection{``Intermediate'' disorder: $g_{A} \rightarrow g_{A}^{\, c}$}\label{IntermediateDisorder}

For larger values of the disorder $g_{A}$, or equivalently, smaller energies with $g^{2} \left|\ln (\omega / \Omega) \right| \sim 1$, Eq.~(\ref{DynamicScaling}) predicts a crossover in the dynamic scaling behavior from Eq.~(\ref{DynamicScalingWeak}) to 
\begin{equation}\label{DynamicScalingIntermediate}
	l \sim \frac{1}{2 g}\sqrt{2  \left| \ln \frac{\omega}{\Omega} \right|} + const.,
\end{equation}
leading to the scaling behaviors
\begin{equation}\label{UEvolveIntermediate}
	U \sim \left(\frac{\Omega}{\omega}\right)^{3} \exp\left[- \frac{1}{g}2\sqrt{2  \left| \ln \frac{\omega}{\Omega} \right|} \right],
\end{equation}
corresponding to the local interaction $\bar{\rho}^{2} + \rho^{2}$, and
\begin{equation}\label{WEvolveIntermediate}
	W \sim \frac{\omega}{\Omega} \exp\left[ 2\sqrt{2  \left| \ln \frac{\omega}{\Omega} \right|}\right],
\end{equation}
corresponding to the local interaction $\bar{\rho} \rho$. Equation (\ref{UEvolveIntermediate}) is 
one of the primary results of our paper.

Equation (\ref{UEvolveIntermediate}) tells us that the same-sublattice four-fermion interaction $\bar{\rho}^{2} + \rho^{2}$ has now become \emph{extremely relevant} towards the low-energy limit $\omega \rightarrow 0$, while the local interaction $\bar{\rho} \rho$ that couples different sublattices together remains irrelevant, albeit enhanced when compared to the clean Dirac point. 
Our result means that the low-energy, delocalized phase of 2D non-interacting 
fermions made possible by the very special chiral sublattice symmetry 
is unstable to same-sublattice (e.g. next-nearest-neighbor) interaction 
effects. On the honeycomb lattice, next-nearest-neighbor interactions 
tend to favor charge-ordering on one triangular sublattice for $U < 0$, 
while for $U > 0$ these interactions are frustrated by the non-bipartite 
nature of the triangular sublattice. 

\begin{figure}
\includegraphics[width=0.4\textwidth]{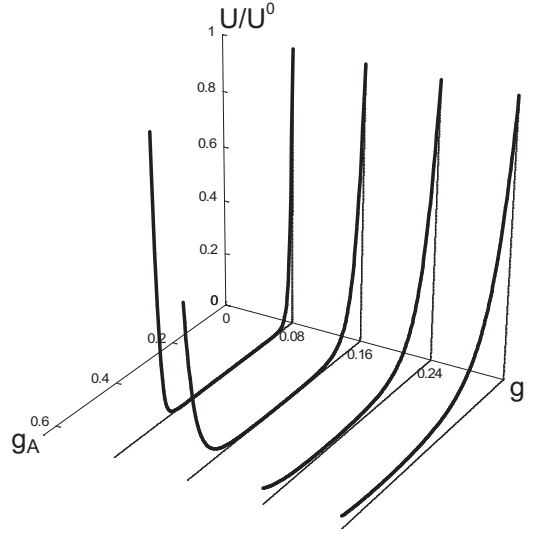}
\caption{Scaling evolution of $U(g_{A})$ for several fixed values of $g$. This figure tracks the irrelevance/relevance of the coupling $U$ associated with the same-sublattice interaction $\bar{\rho}^{2} + \rho^{2}$ as we renormalize to stronger vector potential disorder $g_{A}~\rightarrow~\infty$, or equivalently, smaller energies $\omega~\rightarrow~0$. $U$ is normalized relative to its the bare value $U^{\,0}$ at reference energy scale $\Omega$, and we have tuned $g_{A}^{\,0}~=~0$. The non-interacting, critically delocalized phase resides at $\omega~=~0$ with $g_{A}~=~\infty$. Initially irrelevant near the clean Dirac fixed point ($g_{A}~\ll~1$), $U$ ultimately flows off to strong coupling at larger $g_{A}$, taking the model away from the non-interacting, zero energy delocalized phase.\label{FigUEvolve}}
\end{figure}

Figure \ref{FigUEvolve} shows the RG evolution of the 
same-sublattice interaction strength $U$ with increasing length scale as 
a function of the running random vector potential variance $g_{A}$ 
for several different values of the fixed random mass variance $g$
[see Eq.~(\ref{UofgAEvolve}) below].
This plot describes the following situation: we turn on an infinitesimally 
small interaction strength $U^{\,0}$ and set the bare vector potential 
disorder $g_{A}^{\,0} = 0$ at energy $\omega = \Omega$. For small $g$ 
this places us near the clean Dirac fixed point. Then we run the RG, 
flowing to smaller energies $\omega \rightarrow 0$ and stronger disorder 
$g_{A} \rightarrow \infty$, and we observe the evolution of the 
coupling $U$. When the dynamic scaling behavior crosses over from 
Eq.~(\ref{DynamicScalingWeak}) to Eq.~(\ref{DynamicScalingIntermediate}), 
this interaction strength begins to diverge strongly. 
The analytical expression plotted in Fig.~\ref{FigUEvolve} is given by
\begin{equation}\label{UofgAEvolve}
	U(g_{A}) \sim \exp\left[\frac{3}{2 g^{2}} g_{A}^{2} - \left( \frac{3}{g} + \frac{1}{2g^{2}} \right)g_{A} \right].
\end{equation} 
Equation (\ref{UofgAEvolve}) may be obtained by 
combining Eqs.~(\ref{UEvolve}), (\ref{DynamicScaling}), 
and (\ref{gAEvolve}), setting $g_{A}^{\,0} = 0$. 
The value of $g_{A} \equiv g_{A}^{\, U}$ 
where $U$ becomes relevant 
varies with $g$: for \emph{smaller} values of $g$, the crossover occurs 
at \emph{smaller} values of $g_{A}^{\, U}$. From Eq.~(\ref{UofgAEvolve}), we have
\begin{equation}\label{gAstar}
	g_{A}^{\, U} = \frac{1}{6} + g.
\end{equation}
Note, however, that the transition at small $g$ occurs at much lower energies [since we need $g^{2} \left|\ln (\omega / \Omega) \right|$ to grow from zero ($\omega = \Omega$) to order one ($\omega \sim \Omega \exp(-1/g^{2}) \ll \Omega$)]. 
The freezing transition in the dynamic scaling behavior is 
supposed to occur\cite{MRF} at $g_{A}^{\, c} \sim 1 + \mathcal{O}(g)$ 
for small $g$;
Eq.~(\ref{gAstar}) suggests that for $g \ll 1$, the interaction $U$ becomes relevant 
for $g_{A}$ \emph{well below} $g_{A}^{\, c}$, potentially preventing the RG from ever reaching 
the non-interacting freezing transition.

The instability to interaction effects that we observe here
is perhaps not so surprising when one considers the 
strongly-divergent form of the low-energy density of states in this ``intermediate'' disorder regime given by Eq.~(\ref{DOS1}). The exponent
is
$\alpha = 1/2$ in this equation. We have a picture of a large number of exotic, multifractal states crammed into a very narrow energy window; we might therefore expect that turning on interparticle interactions should produce a very strong effect. The obvious question, then, is why does the same instability \emph{not} occur for weak interactions in the corresponding 1D problem? 

The non-interacting disordered fermion problem with chiral symmetry
[Eq.(\ref{ChiralDef})] {\it in one dimension} may
be formulated as random spin 1/2 quantum $XY$
chain.\cite{SachdevBook}
This random $XY$ chain is asymptotically solvable by a strong 
randomness RG procedure,\cite{DSF,MDH1D}
wherein the spins connected by the strongest random bonds are 
systematically decimated out of the chain, replaced by ever weaker 
bonds connecting the surviving spins. The infrared limit of this RG 
is the so-called random singlet phase\cite{DSF}, 
which
is similar\cite{BOUCHAUD}, in some respects, to the
non-interacting version of the 2D model considered in this 
paper. The addition of weak, homogeneous, 
short-ranged Ising $S_{i}^{z} S_{j}^{z}$ couplings to the random 
$XY$ spin chain, corresponding to density-density interactions in the 
fermion model, does not destabilize this disordered phase; 
in the strong randomness RG, these additional Ising bonds become ever weaker as spins coupled 
with strong $XY$ bonds are removed from the chain. Although some progress\cite{MDH2D} 
has been made in applying these ideas to the non-interacting 
2D chiral Dirac theory, we do not know whether 
this methodology may shed further light on the instability to 
same-sublattice interaction effects that we have found in this paper. 
Presumably, the higher degree of connectedness in the 2D model changes 
the way that the interactions are renormalized relative to the 1D case.

Recall that Eq.~(\ref{UEvolveIntermediate}) only gives the scaling 
behavior of the interaction $\bar{\rho}^{2} + \rho^{2}$ 
(in the intermediately-disordered regime); 
in Eqs.~(\ref{FlowEqg}-\ref{FlowEqsInt}), we 
have not written explicitly the one-loop renormalization effects of 
$U$ upon the disorders strengths $g$ and $g_{A}$ or the 
autorenormalization effects of $U$ upon itself. 
Although we will omit the details here, we have performed the complete 
one-loop RG calculation and found that all of these corrections vanish 
to one loop. (This eliminates the possibility of finding a new, 
interacting critical fixed point to lowest order in $U$.) We also 
know that $U$ cannot renormalize $h$ at one loop, as discussed in the 
paragraph below Eq.~(\ref{BetaFunchAnom}) in Sec.\ \ref{1loop}, 
and thus cannot influence the dynamic scaling behavior given 
by Eq.~(\ref{DynamicScaling}). 

The scaling equations (\ref{UEvolveIntermediate}) 
and (\ref{WEvolveIntermediate}) derived above apply equally to 
the sublattice hopping
model on the square lattice with $\pi$-flux model
per plaquette and real random bond disorder.\cite{HWK}
(The random Hamiltonian is again invariant under
 both chiral and time-reversal symmetries.) 
The operators $\bar{\rho}$ and $\rho$ have the same interpretations as local 
sublattice density operators because their definition [Eq.~(\ref{LocalDensity})] 
relies only upon the definition of the chiral symmetry in the continuum Dirac theory, 
given by Eq.~(\ref{C}). The sublattices of the bipartite square lattice are themselves 
bipartite square lattices, and next-nearest-neighbor interactions tend to produce 
charge-ordering on one sublattice for $U < 0$, and charge stripes for $U > 0$.

\subsubsection{Strong disorder: $g_{A} \gg g_{A}^{\, c}$}
\label{StrongDisorder}

We can imagine that the RG flow reaches successfully the 
regime above the freezing transition $g_{A} > g_{A}^{\, c}$ 
before $U$ grows large enough to 
invalidate our scaling analysis, say by tuning the initial 
interaction strength very small. The ultimate dynamic scaling behavior at 
asymptotically low energies is different from that of Eq.~(\ref{DynamicScaling}); 
instead, for $g_{A} > g_{A}^{\, c}$, the relationship between the RG scale 
$l \rightarrow \infty$ and the bare energy scale $\omega \rightarrow 0$
is given by\cite{MDH2D,MRF}
\begin{equation}\label{DynamicScalingStrong}
	l \sim \left(\frac{1}{c(g)}\left|\ln \frac{\omega}{\Omega} \right|\right)^{\frac{2}{3}},
\end{equation}
with $c(g) \propto g + \mathcal{O}(g^{2})$.
Using
the same logic as above,
this leads
to the scaling behaviors
\begin{equation}\label{UEvolveStrong}
	U \sim \frac{\omega}{\Omega} \exp\left[8 g^{2} \left(\frac{1}{c(g)} \left| \ln \frac{\omega}{\Omega} \right| \right)^{\frac{4}{3}} \right]
\end{equation}
and
\begin{equation}\label{WEvolveStrong}
	W \sim \frac{\omega}{\Omega} \exp\left[4 g \left(\frac{1}{c(g)} \left| \ln \frac{\omega}{\Omega} \right| \right)^{\frac{2}{3}} \right].
\end{equation}
Equation (\ref{UEvolveStrong}) is
another primary result of our paper. 

The main point that we want to stress is that Eq.~(\ref{UEvolveStrong}) 
shows that $U$ is ultimately \emph{even more} relevant beyond the freezing 
transition, so that regardless of whether RG takes us into 
the regime above the freezing transition
before $U$ diverges, we always find that the non-interacting description 
is unstable to the same-sublattice interaction effects characterized by 
the operator $\bar{\rho}^{2} + \rho^{2}$. 
We are able to extend our analysis to this large $g_{A}$ regime because the 
key results leading to Eqs.~(\ref{UEvolveStrong}) and (\ref{WEvolveStrong})
 can be obtained from the $(2+0)$D theory non-perturbatively in $g_{A}$: as
 discussed above Eqs.~(\ref{UEvolve}) and (\ref{WEvolve}), the 1-loop flow 
equations for the disorder couplings $g$ and $g_{A}$ [Eqs.~(\ref{FlowEqg}) 
and (\ref{FlowEqgA})], and for the interaction scaling dimensions [Eq.~(\ref{FlowEqsInt})]
 contain the exact $g_{A}$-dependence (to all orders in $g_{A}$), 
and we combine these results with the non-perturbative dynamic scaling relation 
beyond the freezing transition [Eq.~(\ref{DynamicScalingStrong}), taken from 
Ref.~\onlinecite{MRF}]. This leads to Eqs.~(\ref{UEvolveStrong}) and (\ref{WEvolveStrong}).

\subsection{Summary and outlook}

Within the context of our perturbative one-loop renormalization 
group calculation, we have found that same-sublattice interactions 
become strongly enhanced as one examines the TRI random chiral symmetric Dirac 
model at ever lower energy scales, or equivalently, larger values 
of the running disorder coupling $g_{A}$. We emphasized that 
while the calculation presented in this paper appears perturbative 
in both of the disorder strengths $g$ and $g_{A}$, we can extend 
our results to arbitrarily large values of $g_{A}$ because of 
the known\cite{GLL,MRF} non-perturbative structure of the 
non-interacting (nearly conformal) disordered
Dirac theory.

The validity of our stability analysis is restricted 
to $g \ll 1$; because $g$ does not evolve under the RG in the 
non-interacting phase, we can tune $g$ as small as we wish.  

The same mechanism responsible for the 
instability to interaction effects drives the divergence of the density of states, 
and ultimately the freezing transition, in the non-interacting model. 
We have shown that this instability occurs regardless of whether we reach 
the regime beyond the freezing transition or not.

To lowest order in the interaction strength, we see no signs of the possibility of a new, interacting critical fixed point. We therefore anticipate that the strongly relevant same-sublattice interactions drive the disordered honeycomb model toward some kind of Mott insulating phase. We pointed out that the same picture obtains for the square lattice 
$\pi$-flux model\cite{HWK}.
In the introduction we dubbed the observed evolution of the 
clean
but interacting
Dirac fixed point, \emph{weakly perturbed} by chiral-symmetric disorder $g$ at some finite energy scale $\omega = \Omega$, toward a Mott insulating ground state in the limit of zero energy $\omega \rightarrow 0$, a ``disorder-driven Mott transition.'' In the limit of small $g$, this instability occurs only at very small energies; the scale at which the same-sublattice interactions become relevant occurs \emph{below} the freezing transition for small $g$, and is roughly the same scale
at which the density of states begins to diverge, 
as in Eq.~(\ref{DOS1}) with $\alpha = 1/2$.   

An obvious extension of the present work would be to try to approach 
the interaction effects 
beyond the regime linear in the strengths of the interactions;
indeed, the calculation presented in this paper treats 
the interactions only to lowest order,
addressing only the question of the stability of the 
non-interacting description
to the inclusion of interactions.
The natural framework for such an  
extended investigation is the generalized Finkel'stein 
NL$\sigma$M\cite{BK}, 
which grafts many-particle interactions from the disordered fermion theory directly into Wegner's low energy effective theory of interacting diffusion modes. An advantage of the Finkel'stein NL$\sigma$M is that the perturbative RG is an expansion only in the dimensionless DC resistance; 
therefore, at least formally, even
a one-loop calculation treats the 
interaction couplings to all orders. 
A calculation along these lines
is currently in progress\cite{NewWork}.

\begin{acknowledgments}

We would like to thank Olexei Motrunich and Gil Refael for helpful discussions about the strong randomness RG and the 1D results.
This work was supported in part by NSF under Grant No. DMR-00-75064 and by an NSF Graduate Research Fellowship (MSF).

\end{acknowledgments}

\end{document}